\pdfoutput=1    

\documentclass[%
 superscriptaddress,
 amsmath,amssymb,
 aps,
 pra,
]{revtex4-1}

\usepackage{graphicx}
\usepackage{dcolumn}
\usepackage{bm}

\usepackage{stmaryrd} 
\usepackage{xcolor} 
\usepackage{siunitx}
\usepackage{caption} 
\usepackage{subcaption} 

\graphicspath{{Figures/},.}         

\renewcommand{\vec}[1]{\mathbf{{#1}}}
\newcommand{\pvec}[1]{\mathbf{{#1}}_\parallel }
\newcommand{\vecUnit}[1]{\mathbf{\hat{#1}}}
\newcommand{\pvecUnit}[1]{\mathbf{\hat{#1}}_\parallel }
\newcommand{\xpar}{\pvec{x}}

\newcommand{\qpar}{\pvec{q}}
\newcommand{\imu}{\mathrm{i}}
\newcommand{\dint}[2][]{\!\mathrm{d}^{#1}#2\,}
\newcommand{\dfint}[3][]{\! \frac{\mathrm{d}^{#1}#2}{#3}\,}

\newcommand{\bmath}[1]{\ensuremath{\boldsymbol{#1}}}

\renewcommand{\Im}{\mathrm{Im}\,}

\newcommand{\SF}{\mathcal{F}}
\newcommand{\NSF}{\mathcal{N}}
\newcommand{\KernelD}{K_D}
\newcommand{\KernelN}{K_N}

\begin{document}

\title{The scattering of a scalar beam from isotropic and anisotropic two-dimensional randomly rough Dirichlet or Neumann surfaces: The full angular intensity distributions}

\author{Torstein Storflor Hegge}
\affiliation{Department of Physics, NTNU --- Norwegian University of Science and Technology, NO-7491 Trondheim, Norway}

\author{Torstein Nesse}
\affiliation{Department of Physics, NTNU --- Norwegian University of Science and Technology, NO-7491 Trondheim, Norway}

\author{Alexei A. Maradudin}
\affiliation{Department of Physics and Astronomy, University of California, Irvine, CA 92697, U.S.A.}

\author{Ingve Simonsen}
\email{Ingve.Simonsen@ntnu.no}
\affiliation{Department of Physics, NTNU --- Norwegian University of Science and Technology, NO-7491 Trondheim, Norway}
\affiliation{Department of Petroleum Engineering, University of Stavanger, NO-4036 Stavanger, Norway}
\affiliation{Surface du Verre et Interfaces, UMR 125 CNRS/Saint-Gobain, F-93303 Aubervilliers, France}

\date{\today}

\begin{abstract}
  By the use of Green's second integral identity we determine the field scattered from a two-dimensional randomly rough isotropic or anisotropic Dirichlet or Neumann surface when it is illuminated by a scalar Gaussian beam. The integral equations for the scattering amplitudes are solved nonperturbatively by a rigorous computer simulation approach. The results of these calculations are used to calculate the full angular distribution of the mean differential reflection coefficient. For isotropic surfaces, the results of the present calculations for in-plane scattering are compared with those of earlier studies of this problem. The reflectivities of Dirichlet and Neumann surfaces are calculated as functions of the polar angle of incidence, and the reflectiveties for the two kinds of surfaces of similar roughness parameters are found to be different. For an increasing level of surface anisotropy, we study how the angular intensity distributions of the scattered waves are affected by this level. We find that even small to moderate levels of surface anisotropy can significantly alter the symmetry, shape, and amplitude of the scattered intensity distributions when Gaussian beams are incident on the anisotropic surfaces from different azimuthal angles of incidence. 
\end{abstract}

\maketitle


\section{\label{Sec:Intro}Introduction}
The earliest nonperturbative calculations of the scattering of a field from a two-dimensional randomly rough surface were the studies of the scattering of a scalar beam, incident from vacuum, on a Dirichlet~\cite{Tran1992,Macaskill1993,Tran1993} or a Neumann surface~\cite{Tran1993} carried out by Tran and Maradudin and by Macaskill and Kachoyan. These calculations were based on Green's second integral identity~\cite{*[{}] [{, p.~152.}] Book:Danese1965}. The integral equations for the source functions, namely the values of the field in the vacuum or its normal derivative, evaluated on the rough surface, were transformed into matrix equations which were then solved by iterative approaches. The amplitudes of the scattered field are expressed in terms of these source functions, and the differential reflection coefficient is expressed through the scattering amplitudes. The differential reflection coefficient~(DRC), an experimentally accessible quantity, gives the fraction of the total time-averaged flux incident on the rough surface that is scattered into an element of solid angle about a specified direction of scattering. In scattering from a randomly rough surface it is the average of the DRC over the ensemble of realizations of the surface profile function that is calculated. The result is called the mean differential reflection coefficient~(mean DRC). Multiple scattering effects, in particular enhanced backscattering~\cite{Mendez1987}, were present in the results for the dependence of the mean DRC for in-plane scattering on the polar angle of scattering and a fixed polar angle of incidence.

Although in the years following this pioneering work several nonperturbative  calculations of the scattering of vector fields from impenetrable~\cite{Soriano2001,Simonsen2009-1,Simonsen2010-04,Simonsen2012-05} and penetrable~\cite{Simonsen2009-9,Simonsen2011-03,Simonsen2011-02,Simonsen2011-09,Simonsen2013-05,Simonsen2015-07}  two-dimensional randomly rough surfaces were carried out, little attention seems to have been directed at rigorous nonperturbative calculations of the scattering of incident beams from Dirichlet and Neumann surfaces perhaps because they are simpler than the scattering problems studied in these references. Nevertheless, the results of these calculations are relevant, for example, in ocean acoustics in the context of the scattering of a sonic  wave from a rough ocean floor~\cite{DeSanto1979_note,Ogilvy1987}.

In this paper we revisit the problem of the scattering of a scalar beam from a two-dimensional randomly rough surface, and investigate properties of the scattered field not considered in the earliest studies of this problem~\cite{Tran1992,Macaskill1993,Tran1993}. Thus, in addition to presenting results for scattering from surfaces whose profiles are isotropic Gaussian random processes we also present results for the scattering from surfaces whose profiles are anisotropic Gaussian random processes. In addition to the contribution to the mean differential reflection coefficient from the field scattered incoherently in plane, we also present results for the reflectiveties of these surfaces and the full angular distribution of the intensity of the scattered field. Moreover, these calculations are carried out by means of improved algorithms that yield accurate solutions of the integral equations arising in the scattering theory without the use of iterative methods of the Sturm-Liouville type or modifications thereof~\cite{Tran1992,Macaskill1993,Tran1993}.

\section{\label{Sec:Geometry}Scattering System}

\begin{figure}[tb!]
  \centering
  \includegraphics[width=0.4\textwidth]{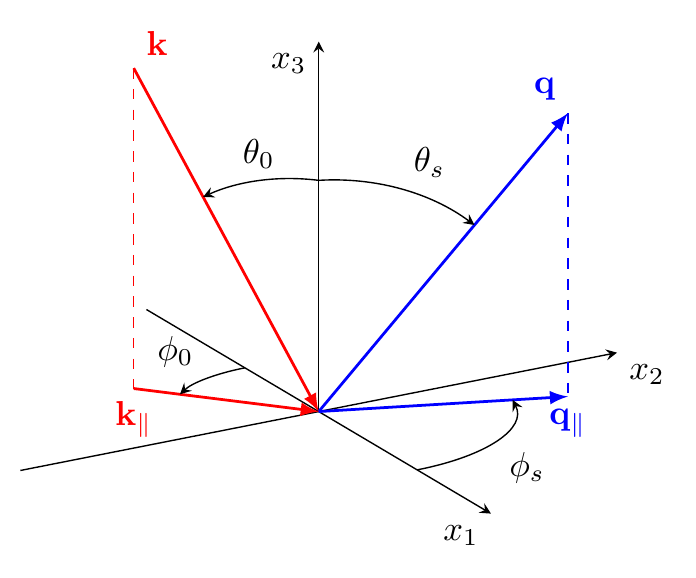}
  \caption{Schematics of the scattering geometry.}
  \label{Fig:ScatteringGeometry}
\end{figure}

The system we consider in this work consists of a medium that supports the propagation of scalar waves without absorption, \textit{e.g.} a liquid, in the region $x_3>\zeta(\pvec{x})$, where $\pvec{x}=(x_1,x_2,0)$ is an arbitrary vector in the plane $x_3=0$, and a medium that is impenetrable to scalar waves in the region $x_3<\zeta(\pvec{x})$~[Fig.~\ref{Fig:ScatteringGeometry}]. The surface profile function $\zeta(\pvec{x})$ is assumed to be a single-valued function of $\pvec{x}$ that is differentiable with respect to $x_1$ and $x_2$, and constitutes a stationary, zero-mean, Gaussian random process. It is defined by
\begin{subequations}
  \label{eq:1}
  \begin{align}
    \label{eq:1a}
    \left< \zeta(\pvec{x}) \right> &= 0
    \\
    \label{eq:1b}
    \left< \zeta(\pvec{x}) \zeta(\pvec{x}') \right> &= \delta^2 W(\pvec{x} - \pvec{x}' ),
\end{align}
\end{subequations}
 where the angle brackets here and in all that follows denote an average over the ensemble of realizations of the surface profile function. The quantity $\delta$, the root-mean-square roughness of the surface, is defined by
\begin{align}
  \label{eq:2}
  \delta &=
           \left< \zeta^2(\pvec{x}) \right>^{\frac{1}{2}}.
\end{align}

The function $W(\pvec{x})$ introduced in Eq.~\eqref{eq:1b} is the \textit{normalized surface height autocorrelation function}, and has the property that, $W(\vec{0})=1$. In what follows we will also require the \textit{power spectrum} of the surface roughness, $g(\pvec{k})$, where $\pvec{k}$ is a two-dimensional wave vector $\pvec{k}=(k_1,k_2,0)$. The power spectrum is the Fourier transform of the normalized surface height auto-correlation function,
\begin{align}
  \label{eq:3}
  g(\pvec{k})
  &=
    \int \dint[2]{\pvec{x}} W(\pvec{x}) \exp\left( -\imu \pvec{k}\cdot \pvec{x} \right).
\end{align}
In this work we will assume the following Gaussian form for $W(\pvec{x})$~\cite{Simonsen2010-04}
\begin{align}
  \label{eq:4}
  W(\pvec{x})
  &=
    \exp \left( -\frac{x_1^2}{a_1^2} -\frac{x_2^2}{a_2^2} \right),
\end{align}
where the characteristic lengths $a_1>0$ and $a_2>0$ are the transverse correlation lengths of the surface roughness along the $x_1$ and $x_2$ axes, respectively.

For the choice of $W(\pvec{x})$ given by Eq.~\eqref{eq:4}, the power spectrum has the Gaussian form
\begin{align}
  \label{eq:5}
  g(\pvec{k})
  &=
    \pi a_1 a_2 \exp\left( -\frac{a_1^2 k_1^2}{4} -\frac{a_2^2 k_2^2}{4} \right).
\end{align}

The surface roughness is said to be \textit{anisotropic} when $a_1\neq a_2$, and is called \textit{isotropic} when $a_1=a_2$. In the latter case the surface height autocorrelation function $W(\pvec{x})$ and the power spectrum $g(\pvec{k})$ depend on $\pvec{x}$ and $\pvec{k}$, respectively, only through their magnitudes $x_\parallel$ and $k_\parallel$ and not on their directions.

\section{\label{sec:theory}Scattering Theory}

The impenetrable surface $x_3=\zeta(\pvec{x})$ is illuminated from the region $x_3>\zeta(\pvec{x})$ by a scalar field of angular frequency $\omega$, that produce a scattered field of the same frequency.  The field $\psi(\vec{x}; t)$ in the region $x_3>\zeta(\pvec{x})$ has the form $\psi(\vec{x}; t) = \psi(\vec{x}|\omega)\exp[-\imu\omega t]$ where the amplitude function $\psi(\vec{x}|\omega)$ is the solution of the  Helmholtz equation
\begin{align}
  \label{eq:6}
  \left[\nabla^2 + \frac{\omega^2}{c^2} \right]  \psi(\vec{x}|\omega)  &= 0,
\end{align}
with $c$ the speed of the field. This field satisfies either (a)~the Dirichlet boundary condition, which corresponds to zero  pressure on the wall, 
\begin{align}
  \label{eq:7}
  \left. \psi(\vec{x}|\omega) \right|_{x_3=\zeta(\pvec{x})} &= 0,
\end{align} 
or (b)~the Neumann boundary condition, which corresponds to zero normal velocity at the wall,
\begin{align}
  \label{eq:8}
  \left.\frac{\partial}{\partial n} \psi(\vec{x}|\omega) \right|_{x_3=\zeta(\pvec{x})} &= 0.
\end{align}
In Eq.~\eqref{eq:8} $\partial/\partial n$ is the derivative along the normal to the surface $x_3=\zeta(\pvec{x})$ at each point of it, directed into the medium of incidence,
\begin{subequations}
  \label{eq:9}
\begin{align}
  \label{eq:9a}
  \frac{\partial}{\partial n} 
  &=
    \frac{1}{ \left[ 1 + \left\{ \bmath{\nabla} \zeta(\pvec{x})\right\}^2  \right]^\frac{1}{2}  }
    \left[
    -\zeta_1(\pvec{x}) \frac{\partial}{\partial x_1}
    -\zeta_2(\pvec{x}) \frac{\partial}{\partial x_2}
    +\frac{\partial}{\partial x_3}
    \right]
  \\
  &\equiv
  \frac{1}{ \left[ 1 + \left\{ \bmath{\nabla} \zeta(\pvec{x})\right\}^2  \right]^\frac{1}{2}  }
  \frac{\partial}{\partial N},
\end{align}
\end{subequations}
where $\zeta_\alpha(\pvec{x})=\partial \zeta(\pvec{x})/\partial x_\alpha$ ($\alpha=1,2$).

To obtain an equation satisfied by $\psi(\vec{x}|\omega)$ that is convenient to solve numerically we begin by introducing the scalar Green's function that satisfies the inhomogeneous Helmholtz equation
\begin{align}
  \label{eq:10}
  \left[\nabla^2 + \frac{\omega^2}{c^2} \right]  g_0(\vec{x}| \vec{x}')
  &=
    -4\pi\, \delta( \vec{x} - \vec{x}' ).
\end{align}
The Green's function has the representations
\begin{subequations}
  \label{eq:11}
  \begin{align}
    \label{eq:11a}
    g_0(\vec{x}| \vec{x}')
    &=
      \frac{
      \displaystyle
      \exp \left[ \imu \frac{\omega}{c} \left| \vec{x} - \vec{x}' \right| \right]
      }{
      \left| \vec{x} - \vec{x}' \right|
      }
    \\
    &=
      \label{eq:11b}
      \int \dfint[2]{q_\parallel}{(2\pi)^2}
      \frac{2\pi \imu}{\alpha_0(q_\parallel)}
      \exp
      \left[
      \imu \pvec{q}\cdot \left( \pvec{x} - \pvec{x}' \right)
      \right]
      \exp
      \left[
      \imu \alpha_0(q_\parallel) \left| x_3 - x_3' \right|
      \right],
  \end{align}
\end{subequations}
where we have introduced the in-plane component of the wave vector $\qpar=(q_1,q_2,0)$ and its corresponding normal component
\begin{align}
  \label{eq:12}
  \alpha_0(q_\parallel)
  &=
    \begin{cases}
      \left[ \frac{\omega^2}{c^2} - q_\parallel^2 \right]^\frac{1}{2}
      & q_\parallel < \frac{\omega}{c}
      \\
      \imu \left[ q_\parallel^2 - \frac{\omega^2}{c^2} \right]^\frac{1}{2}
      & q_\parallel > \frac{\omega}{c}
      \end{cases}.
\end{align}

We next apply Green's second integral identity~\cite[p.~152]{Book:Danese1965}
\begin{align}
  \label{eq:13} 
  \int\limits_\Omega \dint[3]{x_\parallel}
  \left( u \nabla^2v - v \nabla^2  u \right)
  &=
    \int\limits_{\Sigma} \dint[]{s}     
    \left( u \frac{\partial v}{ \partial \nu} - v \frac{\partial u}{ \partial \nu }  \right),
\end{align}
where $u(\vec{x})$ and $v(\vec{x})$ are arbitrary scalar functions of $\vec{x}$ defined in a volume $\Omega$ that is bounded by a closed surface $\Sigma$. The derivative $\partial /\partial\nu$ is taken along the normal to the surface $\Sigma$ at each point of it, directed away from the volume $\Omega$.

We assume that the volume $\Omega$ is the region $x_3>\zeta(\pvec{x})$, while the surface $\Sigma$ is the union of  the rough surface $x_3=\zeta(\pvec{x})$, which we denote by $S$, and a hemispherical cap of infinite radius in the upper half space, which we denote by $S^{(\infty)}$.  Then, on setting $u=\psi(\vec{x}|\omega)$ and $v=g_0(\vec{x}|\vec{x}')$ in Eq.~\eqref{eq:13}, and taking into account Eqs.~\eqref{eq:6} and \eqref{eq:10}, we obtain
\begin{align}
  \label{eq:14}
  -4\pi\; \theta\left(x_3' - \zeta(\pvec{x}') \right) \psi( \vec{x}'|\omega )
  &=
    -\int\limits_S \dint[]{S}
    \left[
    \psi(\vec{x}|\omega) \frac{\partial }{\partial n} g_0(\vec{x}|\vec{x}')
    -
    g_0(\vec{x}|\vec{x}') \frac{\partial }{\partial n} \psi(\vec{x}|\omega) 
    \right]
    \notag \\
  & \quad
    +
    \int\limits_{S^{(\infty)}} \dint[]{S}
    \left[
    \psi(\vec{x}|\omega) \frac{\partial }{\partial \nu} g_0(\vec{x}|\vec{x}')
    -
    g_0(\vec{x}|\vec{x}') \frac{\partial }{\partial \nu} \psi(\vec{x}|\omega) 
    \right],
\end{align}
where $\theta(z)$ is the Heaviside unit step function. Because the  scattered field satisfies a radiation condition at infinity, its contribution to the surface integral over the hemispherical cap of infinite radius $S^{(\infty)}$ on the right-hand side of Eq.~\eqref{eq:14} vanishes. This integral therefore yields $-4\pi\psi(\vec{x}|\omega)_\mathrm{inc}$ where  $\psi(\vec{x}|\omega)_\mathrm{inc}$ is the incident field. By using the symmetry of $g_0(\vec{x}|\vec{x}')$, we can now rewrite Eq.~\eqref{eq:14} as
\begin{align}
  \label{eq:15}
  \theta\left(x_3 - \zeta(\pvec{x}) \right) \psi( \vec{x}|\omega )
  &=
    \psi(\vec{x}|\omega)_\mathrm{inc}
    +
    \frac{1}{4\pi}
    \int\limits_{S} \dint[]{S'}
    \left[
    \left( \frac{\partial }{\partial n'} g_0(\vec{x}|\vec{x}') \right)
    \psi(\vec{x}'|\omega) 
    -
    g_0(\vec{x}|\vec{x}') \frac{\partial }{\partial n'} \psi(\vec{x}'|\omega) 
    \right].    
\end{align}
Since we have assumed that the surface profile function is a single valued function of $\pvec{x}$, we can replace the integration over the surface $S$ by integration over the plane $x_3=0$ with the use of the relation $\dint{}{S}=\left[ 1 + \left\{ \bmath{\nabla} \zeta(\pvec{x})\right\}^2  \right]^\frac{1}{2}  \dint[2]{x_\parallel}$,
\begin{align}
  \label{eq:16}
  \theta\left(x_3 - \zeta(\pvec{x}) \right) \psi( \vec{x}|\omega )
  &=
    \psi(\vec{x}|\omega)_\mathrm{inc}
    +
    \frac{1}{4\pi}
    \int\dint[2]{x_\parallel'}
    \left\{
    \left. \left[
    \frac{\partial }{\partial N'} g_0(\vec{x}|\vec{x}')
    \right]\right|_{x_3'=\zeta(\pvec{x}')}
    \SF(\pvec{x}'|\omega) 
    -
    \left. \left[ g_0(\vec{x}|\vec{x}') \right] \right|_{x_3'=\zeta(\pvec{x}')}
    \NSF(\pvec{x}'|\omega)
    \right\},    
\end{align}
where
\begin{subequations}
  \label{eq:17}
  \begin{align}
    \label{eq:17a}
    \SF(\pvec{x}|\omega)
    &=
      \left. \psi(\vec{x}|\omega) \right|_{x_3=\zeta(\pvec{x})} 
    \\
    \label{eq:17b}
    \NSF(\pvec{x}|\omega)
    &=
      \left.\frac{\partial}{\partial N} \psi(\vec{x}|\omega) \right|_{x_3=\zeta(\pvec{x})}.
  \end{align}
\end{subequations}
With the use of either the Dirichlet or Neumann boundary condition, we can simplify Eq.~\eqref{eq:16}.

\subsection{Dirichlet boundary condition}
The Dirichlet boundary condition, Eq.~\eqref{eq:7}, can be called an acoustically soft-wall boundary condition. Its assumption simplifies Eq.~\eqref{eq:16} to
\begin{align}
  \label{eq:18}
  \theta\left(x_3 - \zeta(\pvec{x}) \right) \psi( \vec{x}|\omega )
  &=
    \psi(\vec{x}|\omega)_\mathrm{inc}
    -
    \frac{1}{4\pi}
    \int\dint[2]{x_\parallel'}
    \left. \left[ g_0(\vec{x}|\vec{x}') \right] \right|_{x_3'=\zeta(\pvec{x}')}
    \NSF(\pvec{x}'|\omega).
\end{align}
The scattered field is the second term on the right-hand side of Eq.~\eqref{eq:18}. With the use of the representation for $g_0(\vec{x}|\vec{x}')$ given by Eq.~\eqref{eq:11b}, the scattered field can be written as
\begin{align}
  \label{eq:19}
  \psi(\vec{x}|\omega)_{\mathrm{sc}}
  &=
    \int \dfint[2]{q_\parallel}{(2\pi)^2}
    R_D(\pvec{q}, \omega)
    \exp
    \left[
    \imu \pvec{q}\cdot \pvec{x}
    +
    \imu \alpha_0(q_\parallel) x_3
    \right],
\end{align}
where
\begin{align}
  \label{eq:20}
  R_D(\pvec{q}, \omega)
  &=
    -\frac{\imu}{2\alpha_0(q_\parallel)}
    \int \dint[2]{x_\parallel}
    \NSF(\pvec{x}|\omega)
    \exp
    \left[
    -
    \imu \pvec{q}\cdot \pvec{x}
    -
    \imu \alpha_0(q_\parallel) \zeta(\pvec{x})    
    \right].
\end{align}

\subsection{Neumann boundary condition}

The Neumann boundary condition, Eq.~\eqref{eq:8}, can be called an acoustically  hard-wall boundary condition. Its use simplifies Eq.~\eqref{eq:16} to
\begin{align}
  \label{eq:21}
  \theta\left(x_3 - \zeta(\pvec{x}) \right) \psi( \vec{x}|\omega )
  &=
    \psi(\vec{x}|\omega)_\mathrm{inc}
    +
    \frac{1}{4\pi}
    \int\dint[2]{x_\parallel'}
    \left. \left[
    \frac{\partial }{\partial N'} g_0(\vec{x}|\vec{x}')
    \right]\right|_{x_3'=\zeta(\pvec{x}')}
    \SF(\pvec{x}'|\omega).
\end{align}
The scattered field is the second  term on the right-hand side of this equation. With the use of the representation of $g_0(\vec{x}|\vec{x}')$ given by Eq.~\eqref{eq:11b}, it assumes the form
\begin{align}
  \label{eq:22}
  \psi(\vec{x}|\omega)_{\mathrm{sc}}  
  &=
    \int \dfint[2]{q_\parallel}{(2\pi)^2}
    R_N(\pvec{q}, \omega)
    \exp
    \left[
    \imu \pvec{q}\cdot \pvec{x}
    +
    \imu \alpha_0(q_\parallel) x_3
    \right],
\end{align}
where
\begin{align}
  \label{eq:23}
  R_N(\pvec{q}, \omega)
  &=
    -\frac{1}{2\alpha_0(q_\parallel)}
    \int \dint[2]{x_\parallel}
    \SF(\pvec{x}|\omega)
    \left[
    \pvec{q} \cdot \bmath{\nabla} \zeta(\pvec{x})
    -
    \alpha_0(q_\parallel)     
    \right]
    \exp
    \left[
    -
    \imu \pvec{q}\cdot \pvec{x}
    -
    \imu \alpha_0(q_\parallel) \zeta(\pvec{x})    
    \right].
\end{align}

\section{\label{Sec:meanDRC}The mean differential reflection coefficient}

The scattering amplitudes $R_D(\pvec{q},\omega)$ and $R_N(\pvec{q},\omega)$ play a central role in the theory of the scattering of a wave from a randomly rough surface. This is because the coherent (specular) and incoherent (diffuse) scattering of the wave can be described quantitatively in terms of these amplitudes, through their presence in the expressions for the mean differential reflection coefficient.

The differential reflection coefficient $\partial R/\partial\Omega_s$ is defined such that $(\partial R/\partial\Omega_s)\mathrm{d}\Omega_s$ is the fraction of the total time-averaged flux incident on the surface that is scattered into the element of solid angle $\mathrm{d}\Omega_s$ about the direction of scattering defined by the polar and azimuthal angles of scattering $\theta_s$ and $\phi_s$, respectively. 

Because in numerical simulations of scattering from  a rough surface only a finite region of the plane $x_3=0$ can be covered by a two-dimensional rough surface, we assume for the incident field a Gaussian beam that illuminates only that finite region of the $x_3$ plane, minimizing edge effects in the scattering thereby. The Gaussian incident field is given by
\begin{align}
  \label{eq:24}
  \psi( \vec{x} | \omega)_{\mathrm{inc}}
  &=
    \frac{w^2}{2\pi}
    \int\limits_{q_\parallel < \frac{\omega}{c}} \dint[2]{q_\parallel}
    \exp
    \left[
    \imu \pvec{q} \cdot \pvec{x} - \imu \alpha_0(q_\parallel)x_3
    \right]
    \exp
    \left[
    - \frac{w^2}{2} \left( \pvec{q} - \pvec{k} \right)^2
    \right].
\end{align}
The magnitude of the total time-averaged flux incident on the surface is given by 
\begin{align}
  \label{eq:25}
  P_{\mathrm{inc}} 
  &=  
    -
    A\,
    \Im\!\!
    \int\limits_{-\frac{L_1}{2}}^{\frac{L_1}{2}} \!\! \dint{x_1}
    \int\limits_{-\frac{L_2}{2}}^{\frac{L_2}{2}} \!\! \dint{x_2}
    \psi^*(\vec{x}|\omega)_{\mathrm{inc}}
    \frac{\partial \psi(\vec{x}|\omega)_{\mathrm{inc}}}{\partial x_3},
\end{align}
where $L_1$ and $L_2$ are the lengths of the scattering surface along the $x_1$ and $x_2$ axes, respectively, while $A$ is a coefficient that drops out of the expression for the differential reflection coefficient~\cite{*[{}] [{, pp.~918--920.} ] Book:Born2002}. The minus sign that appears on the right-hand side of Eq.~\eqref{eq:25} compensates for the fact that the incident flux is negative. On substituting Eq.~\eqref{eq:24} into Eq.~\eqref{eq:25}, we obtain
\begin{align}
  \label{eq:26}
  P_{\mathrm{inc}}
  &=
    A \left( \frac{w^2}{2\pi}\right)^2
    \Im \; \imu \!
    \int\limits_{q_\parallel < \frac{\omega}{c}} \dint[2]{q_\parallel}
    \int\limits_{q_\parallel' < \frac{\omega}{c}} \dint[2]{q_\parallel'}
    \alpha_0(q_\parallel')
    \exp
    \left\{
    \imu \left[\alpha_0^*(q_\parallel) - \alpha_0(q_\parallel') \right] x_3
    \right\}
    \notag
  \\
  &
    \qquad \times 
    \exp
    \left[
    -\frac{w^2}{2} \left( \pvec{q} - \pvec{k}  \right)^2
    -\frac{w^2}{2} \left( \pvec{q'}- \pvec{k}  \right)^2
    \right]
    %
  %
    \int\limits_{-\frac{L_1}{2}}^{\frac{L_1}{2}} \!\! \dint{x_1}
   \int\limits_{-\frac{L_2}{2}}^{\frac{L_2}{2}} \!\! \dint{x_2}
    \exp
    \left[
    - \imu
    \left(
    \pvec{q} - \pvec{q}'
    \right)
    \cdot \pvec{x}
    \right]
    \notag
  \\
  &=
    A w^4
    \int\limits_{q_\parallel < \frac{\omega}{c}} \dint[2]{q_\parallel}
    \alpha_0(q_\parallel)
    \exp \left[ -w^2 \left( \pvec{q} - \pvec{k}   \right)^2 \right],
\end{align}
in the limit as $L_{1,2}\rightarrow \infty$. By carrying out the angular integration in Eq.~\eqref{eq:26} we obtain an expression for $  P_{\mathrm{inc}}$ as a one-dimensional integral
\begin{align}
  \label{eq:27}
  P_{\mathrm{inc}}
  &=
    2\pi A w^4 \exp \left( -w^2 k_\parallel^2 \right)
    \int\limits_{0}^{\frac{\omega}{c}} \dint{q_\parallel}
    q_\parallel\alpha_0(q_\parallel)
    \operatorname{I}_0\!\left( 2 w^2 q_\parallel k_\parallel \right)
    \exp \left( -w^2 q_\parallel^2 \right),
\end{align}
where $\operatorname{I}_0(z)$ is the modified Bessel function of the first kind and order zero. An alternative expression for $P_{\mathrm{inc}}$ is obtained when we make the  change of variable $q_\parallel = (\omega/c)\sin\vartheta$, namely
\begin{align}
  \label{eq:28}
  P_{\mathrm{inc}}
  &=
    2\pi A w^4
    \left( \frac{\omega}{c} \right)^3
    \exp \left( -w^2 k_\parallel^2 \right)
    \int\limits_{0}^{\frac{\pi}{2}} \dint{\vartheta}
    \sin\vartheta \cos^2\vartheta \,
    \operatorname{I}_0\!\left( 2 w^2 \frac{\omega}{c} k_\parallel \sin\vartheta \right)
    \exp \left[ -\left(\frac{w \omega}{c}\right)^2\sin^2\vartheta \right].
\end{align}
The relation between the wave vector $\pvec{k}$  and the polar and azimuthal angles of incidence, $\theta_0$ and $\phi_0$, respectively, is
\begin{align}
  \label{eq:29}
  \pvec{k} = \frac{\omega}{c} \sin\theta_0 \left( \cos\phi_0, \sin\phi_0, 0 \right),
\end{align}
so that $k_\parallel =(\omega/c)\sin\theta_0$. With this result the expression for $P_{\mathrm{inc}}$, Eq.~\eqref{eq:28}, becomes
\begin{subequations}
  \label{eq:30}
\begin{align}
  \label{eq:30a}
  P_{\mathrm{inc}}
  &=
    P_{\mathrm{inc}}(\theta_0)
    =
    A\, p_{\mathrm{inc}}(\theta_0),
\end{align}
where
\begin{align}
  \label{eq:30b}
  p_{\mathrm{inc}}(\theta_0)
  &=
    2\pi w^4 \left(\frac{\omega}{c}\right)^3
    \exp
    \left[
    - \left( \frac{w\omega}{c} \right)^2 \sin^2\theta_0
    \right]
    \int\limits_{0}^{\frac{\pi}{2}} \dint{\vartheta}
    \sin\vartheta \cos^2\!\vartheta \,
    \operatorname{I}_0\!\left( 2 \left(\frac{w\omega}{c}\right)^2 \sin\theta_0 \sin\vartheta \right)
    \exp \left[ -\left( \frac{w\omega}{c}\right)^2 \sin^2\vartheta \right].    
\end{align}
\end{subequations}
The expressions for $P_{\mathrm{inc}}$ given by Eqs.~\eqref{eq:27} and \eqref{eq:30} have to be evaluated numerically.

The field scattered from either a Dirichlet or Neumann surface can be  written as (see Eqs.~\eqref{eq:19} and \eqref{eq:22})
\begin{align}
  \label{eq:31}
  \psi( \vec{x} | \omega )_{\mathrm{sc}}
  &=
    \int\dfint[2]{q_\parallel}{(2\pi)^2}
    R(\pvec{q},\omega)
    \exp
    \left[
    \imu \pvec{q} \cdot \pvec{x}
    +
    \imu \alpha_0(q_\parallel) x_3
    \right],
\end{align}
where $R(\pvec{q},\omega)$ is either $R_D(\pvec{q},\omega)$ or $R_N(\pvec{q},\omega)$. The total time-averaged scattered flux is given by 
\begin{align}
  \label{eq:32}
  P_{\textrm{sc}} 
  &=  
    A \;
    \Im\!\!
    \int\limits_{-\frac{L_1}{2}}^{\frac{L_1}{2}} \!\! \dint{x_1}
    \int\limits_{-\frac{L_2}{2}}^{\frac{L_2}{2}} \!\! \dint{x_2}
    \psi^*(\vec{x}|\omega)_{\textrm{sc}}
    \frac{\partial \psi(\vec{x}|\omega)_{\textrm{sc}}}{\partial x_3}.
\end{align}
When we substitute Eq.~\eqref{eq:31} into Eq.~\eqref{eq:32} the latter becomes
\begin{align}
  \label{eq:33}
  P_{\textrm{sc}} 
  &=  
    A \;
    \Im\!\!
    \int \dfint[2]{q_\parallel}{(2\pi)^2}
    \int \dfint[2]{q'_\parallel}{(2\pi)^2}
    \imu \alpha_0(q'_\parallel)
    R^*(\pvec{q},\omega) R(\pvec{q}',\omega)
    \exp
    \left\{
    -\imu \big[
    \alpha_0^*(q_\parallel) - \alpha_0(q'_\parallel) 
    \big] x_3
    \right\}
    \notag
  \\
  & 
    \qquad \qquad \qquad \qquad
    \qquad
    \times
    \int\limits_{-\frac{L_1}{2}}^{\frac{L_1}{2}} \!\! \dint{x_1}
    \int\limits_{-\frac{L_2}{2}}^{\frac{L_2}{2}} \!\! \dint{x_2}
    \exp\left[ -\imu\big(\pvec{q}-\pvec{q}'\big)\cdot \pvec{x}\right]
    \notag
  \\
  &=
    A\; \Im \imu \!\!
    \int \dfint[2]{q_\parallel}{(2\pi)^2}
    \alpha_0(q_\parallel)
    \left| R(\pvec{q},\omega) \right|^2
    \exp \left[ -2\Im\alpha_0(q_\parallel) x_3 \right],
\end{align}
in the limit where $L_{1,2}\rightarrow \infty$.  The function $\alpha_0(q_\parallel)$  is real for $0< q_\parallel<\omega/c$, and imaginary for $q_\parallel>\omega/c$. Thus we finally obtain for $  P_{\textrm{sc}}$ the result
\begin{align}
  \label{eq:34}
  P_{\textrm{sc}} 
  &=
    A
    \int\limits_{q_\parallel<\omega/c}
    \dfint[2]{q_\parallel}{(2\pi)^2}
    \alpha_0(q_\parallel)
    \left| R(\pvec{q}, \omega) \right|^2.    
\end{align}

The relation between the wave vector $\pvec{q}$ and the polar and azimuthal angles of scattering $\theta_s$ and $\phi_s$ is
\begin{align}
  \label{eq:35}
  \pvec{q} = \frac{\omega}{c} \sin\theta_s \left( \cos\phi_s, \sin\phi_s, 0 \right),
\end{align}
so that
\begin{subequations}
  \begin{align}
    q_\parallel &= \frac{\omega}{c} \sin\theta_s
    \\
    \alpha_0(q_\parallel) &= \frac{\omega}{c} \cos\theta_s,
  \end{align}
  while
  \begin{align}
    \dint[2]{q_\parallel}
    &=
      \left(\frac{\omega}{c}\right)^2 \sin\theta_s\, \mathrm{d}{\Omega_s}, 
  \end{align}
\end{subequations}
where $\mathrm{d}{\Omega_s}=\sin\theta_s\,\mathrm{d}{\theta_s} \mathrm{d}{\phi_s}$ is the element of solid angle at $(\theta_s,\phi_s)$.
The total time-averaged scattered flux can then be  written in the form 
\begin{align}
  \label{eq:37}
  P_{\textrm{sc}} 
  &=
    \int \dint{\Omega_s}
    P_{\mathrm{sc}} (\theta_s, \phi_s),
\end{align}
where
\begin{align}
  \label{eq:38}
  P_{\mathrm{sc}} (\theta_s, \phi_s)
  &=
    A \left( \frac{\omega}{2\pi c}\right)^2 \frac{\omega}{c} \cos^2\theta_s
    \left| R(\pvec{q}, \omega ) \right|^2.        
\end{align}

By definition the differential reflection coefficient is given by 
\begin{align}
  \label{eq:39}
  \frac{\partial R }{ \partial \Omega_s}
  &=
    \frac{ P_{\mathrm{sc}} (\theta_s, \phi_s) }{P_{\mathrm{inc}} (\theta_0)}
   =   
   \frac{1}{4\pi^2} \left( \frac{\omega}{c}\right)^3 
    \frac{ \cos^2\theta_s }{ p_{\mathrm{inc}} (\theta_0) }   
    \left| R(\pvec{q},\omega) \right|^2.        
\end{align}

Since we are considering scattering from a randomly rough surface, it is not the differential reflection coefficient itself that we need to calculate, but rather its average over the ensemble of realizations of the surface profile function. The resulting mean differential reflection coefficient is given by
\begin{align}
  \label{eq:40}
  \left< \frac{\partial R }{ \partial \Omega_s} \right>
  &=
    \frac{1}{4\pi^2} \left( \frac{\omega}{c}\right)^3 
    \frac{ \cos^2\theta_s }{ p_{\mathrm{inc}} (\theta_0) }          
    \left< \left| R(\pvec{q},\omega) \right|^2 \right>.            
\end{align}

If we write the scattering amplitude $R(\pvec{q}|\pvec{k})$ as the sum of its mean value and of its fluctuation away from the mean value,
\begin{align}
  \label{eq:41}
  R(\pvec{q}|\pvec{k})
  &=
    \left< R(\pvec{q}|\pvec{k}) \right>
    +
    \left[
    R(\pvec{q}|\pvec{k}) - \left< R(\pvec{q}|\pvec{k}) \right>
    \right],
\end{align}
we find that each term contributes separately to the mean differential reflection coefficient, which then takes the form
\begin{align}
  \label{eq:42}
  \left< \frac{\partial R }{ \partial \Omega_s} \right>
  &=
    \left< \frac{\partial R }{ \partial \Omega_s} \right>_{\mathrm{coh}}
    +
    \left< \frac{\partial R }{ \partial \Omega_s} \right>_{\mathrm{incoh}},
\end{align}
where
\begin{align}
  \label{eq:43}
  \left< \frac{\partial R }{ \partial \Omega_s} \right>_{\mathrm{coh}}
  &=
    \frac{1}{4\pi^2} \left( \frac{\omega}{c}\right)^3 
    \frac{ \cos^2\theta_s }{ p_{\mathrm{inc}} (\theta_0) } 
    \left| \left< R(\pvec{q},\omega) \right> \right|^2
\end{align}
and
\begin{align}
  \label{eq:44}
  \left< \frac{\partial R }{ \partial \Omega_s} \right>_{\mathrm{incoh}}
  &=
    \frac{1}{4\pi^2} \left( \frac{\omega}{c}\right)^3 
    \frac{ \cos^2\theta_s }{ p_{\mathrm{inc}} (\theta_0) } 
    \left[
    \Big< \Big| R(\pvec{q},\omega) \Big|^2\Big>
    -
    \left| \Big< R(\pvec{q},\omega)  \Big> \right|^2
    \right].
\end{align}
The former term gives the contribution to the mean differential reflection coefficient from the coherently (specularly) scattered field, while the latter term gives the contribution from the incoherently (diffusely) scattered field.

The reflectivity of the randomly rough surface is given by
\begin{align}
  \label{eq:45}
  \mathcal{R}(\theta_0,\phi_0)
  &=
    \int\limits_0^{\frac{\pi}{2}} \dint{\theta_s} \sin\theta_s
    \int\limits_{-\pi}^{\pi} \dint{\phi_s}
    \left< \frac{\partial R }{ \partial \Omega_s} \right>_{\mathrm{coh}}.    
\end{align}
The dependence of the reflectivity on the azimuthal angle of incidence $\phi_0$ arises only when the randomly rough  surface is defined by a surface profile fucntion  that is a stationary  anisotropic  random process. When the surface profile function is a stationary isotropic random process, the reflectivity is independent of $\phi_0$.

\section{\label{Sec:SrcFunc}Equations Satisfied by the Source Functions}

We see from Eqs.~\eqref{eq:18} and \eqref{eq:21} that once the source functions $\SF(\pvec{x}|\omega)$ and $\NSF(\pvec{x}|\omega)$ are known these equations allow the scattered fields to be determined at any point in the region $x_3>\zeta(\pvec{x})$. To obtain the equations satisfied by these functions we consider the cases of Dirichlet and Neumann surfaces in turn.  

\subsection{Dirichlet Surfaces}
To obtain the equation for the source function $\NSF(\pvec{x}|\omega)$ we first assume that $x_3>\zeta(\pvec{x})$ and apply the derivative operator $\partial/\partial N$ to both sides of Eq.~\eqref{eq:18} to obtain
\begin{align}
  \label{eq:46}
  \theta\left(x_3 - \zeta(\pvec{x}) \right) \frac{\partial}{\partial N} \psi( \vec{x}|\omega )
  &=
    \frac{\partial}{\partial N} 
    \psi(\vec{x}|\omega)_\mathrm{inc}
    -
    \frac{1}{4\pi}
    \int\dint[2]{x_\parallel'}
    \left. \left[
    \frac{\partial}{\partial N} g_0(\vec{x}|\vec{x}')
    \right] \right|_{x_3'=\zeta(\pvec{x}')}
    \NSF(\pvec{x}'|\omega).
\end{align}
We next evaluate this equation at $x_3=\zeta(\pvec{x})+\eta$ and at $x_3=\zeta(\pvec{x})-\eta$, where $\eta$ is a positive infinitesimal, add the resulting equations, and let $\eta\rightarrow 0$. The result is
\begin{align}
  \label{eq:47}
  \NSF(\pvec{x}|\omega)
  &=
    2 \NSF(\pvec{x}|\omega)_{\mathrm{inc}}
    -
    \frac{1}{4\pi}
    \int\dint[2]{x_\parallel'}
    \left\{
    \left.
    \left[
    \frac{\partial}{\partial N} g_0(\vec{x}|\vec{x}')
    \right] \right|_{ \substack{ x_3=\zeta(\pvec{x})+\eta \\x_3'=\zeta(\pvec{x}')\quad } }
    +
    \left.
    \left[
    \frac{\partial}{\partial N} g_0(\vec{x}|\vec{x}')
    \right] \right|_{ \substack{ x_3=\zeta(\pvec{x})-\eta \\ x_3'=\zeta(\pvec{x}')\quad  } }
    \right\}
    \NSF(\pvec{x}'|\omega),
\end{align}
where $\NSF(\pvec{x}|\omega)_{\mathrm{inc}}=[\partial \psi(\vec{x}|\omega)_\mathrm{inc} /\partial N]\big|_{x_3=\zeta(\pvec{x})}$. With the use of the results 
\begin{subequations}
  \label{eq:48}
\begin{align}
  \label{eq:48a}
  \left.
  \left[
  \frac{\partial}{\partial N} g_0(\vec{x}|\vec{x}')
  \right] \right|_{ \substack{ x_3=\zeta(\pvec{x})+\eta \\x_3'=\zeta(\pvec{x}')\quad } }
  +
  \left.
  \left[
  \frac{\partial}{\partial N} g_0(\vec{x}|\vec{x}')
  \right] \right|_{ \substack{ x_3=\zeta(\pvec{x})-\eta \\ x_3'=\zeta(\pvec{x}')\quad  } }
&=
  2 \mathcal{P}
   \left\llbracket
  \frac{\partial}{\partial N} g_0(\vec{x}|\vec{x}')
   \right\rrbracket,
\end{align}
where $\mathcal{P}$ denotes the Cauchy principle value, and
\begin{align}
  \label{eq:48b}
  \big\llbracket f(\vec{x}|\vec{x}') \big\rrbracket
  &=
    f(\vec{x} |\vec{x}')\Big|_{\substack{x_3=\zeta(\xpar)\\x_3'=\zeta(\xpar')}},
\end{align}
\end{subequations}
we obtain finally the equation satisfied by $\NSF(\pvec{x}|\omega)$
\begin{subequations}
    \label{eq:49}
\begin{align}
  \label{eq:49a}
  \NSF(\pvec{x}|\omega)
  &=
    2 \NSF(\pvec{x}|\omega)_{ \mathrm{inc} }
    -
    \frac{1}{2\pi} \mathcal{P}
    \int \dint[2]{x_\parallel'}
    \KernelD(\pvec{x} | \pvec{x}' )
    \NSF(\pvec{x}'|\omega),
\end{align}
where we have defined the (Dirichlet) kernel 
\begin{align}
  \label{eq:49b}
  \KernelD(\pvec{x} | \pvec{x}' )
  &=
    \left\llbracket
    \frac{\partial}{\partial N} g_0(\vec{x}|\vec{x}')
    \right\rrbracket.
\end{align}
\end{subequations}

\subsection{Neumann Surfaces}

To obtain the equation satisfied by the source function $\SF(\pvec{x}|\omega)$, we begin by evaluating Eq.~\eqref{eq:21} at $x_3=\zeta(\pvec{x})+\eta$ and $x_3=\zeta(\pvec{x})-\eta$, adding the resulting equations and then letting $\eta\rightarrow 0$. The result is
\begin{align}
  \label{eq:50}
  \SF(\pvec{x} | \omega)
  &=
    2   \SF(\pvec{x} | \omega)_{\mathrm{inc}}
    +
    \frac{1}{4\pi}
    \int\dint[2]{x_\parallel'}
    \left\{
    \left.
    \left[
    \frac{\partial}{\partial N'} g_0(\vec{x}|\vec{x}')
    \right] \right|_{ \substack{ x_3=\zeta(\pvec{x})+\eta \\x_3'=\zeta(\pvec{x}')\quad } }
    +
    \left.
    \left[
    \frac{\partial}{\partial N'} g_0(\vec{x}|\vec{x}')
    \right] \right|_{ \substack{ x_3=\zeta(\pvec{x})-\eta \\ x_3'=\zeta(\pvec{x}')\quad  } }
    \right\}
    \SF(\pvec{x}'|\omega),    
\end{align}
where $\SF(\pvec{x}|\omega)_{\mathrm{inc}}=\psi(\vec{x}|\omega)_\mathrm{inc}\big|_{x_3=\zeta(\pvec{x})}$. We next use the result
\begin{align}
  \label{eq:51}
  \left.
  \left[
  \frac{\partial}{\partial N'} g_0(\vec{x}|\vec{x}')
  \right] \right|_{ \substack{ x_3=\zeta(\pvec{x})+\eta \\x_3'=\zeta(\pvec{x}')\quad } }
  +
  \left.
  \left[
  \frac{\partial}{\partial N'} g_0(\vec{x}|\vec{x}')
  \right] \right|_{ \substack{ x_3=\zeta(\pvec{x})-\eta \\ x_3'=\zeta(\pvec{x}')\quad  } }
&=
  2 \mathcal{P}
   \left\llbracket
  \frac{\partial}{\partial N'} g_0(\vec{x}|\vec{x}')
   \right\rrbracket,
\end{align}
to obtain finally the equation satisfied by the source function $\SF(\pvec{x}|\omega)$
\begin{subequations}
  \label{eq:52}
\begin{align}
  \label{eq:52a}
  \SF(\pvec{x} | \omega)
  &=
    2 \SF(\pvec{x} | \omega)_{\mathrm{inc}}
    +
    \frac{1}{2\pi} \mathcal{P}
    \int \dint[2]{x_\parallel'}
    \KernelN(\pvec{x} | \pvec{x}' )
    \SF(\pvec{x}'|\omega),
\end{align}
where the (Neumann) kernel is 
\begin{align}
  \label{eq:52b}
  \KernelN(\pvec{x} | \pvec{x}' )
  &=
    \left\llbracket
    \frac{\partial}{\partial N'} g_0(\vec{x}|\vec{x}')
    \right\rrbracket.    
\end{align}
\end{subequations}

\section{\label{sec:NumSol}Numerical Solution of the Equations for the Source Functions}

The integral equations~\eqref{eq:49} and \eqref{eq:52} satisfied by the source functions $\NSF(\pvec{x}|\omega)$ and $\SF(\pvec{x}|\omega)$, respectively, have to be solved numerically. In this section we show how this is done in scattering from a Dirichlet and Neumann surface. 

To solve Eq.~\eqref{eq:49} we first replace  integration over the entire $x_1'x_2'$ plane by integration over the finite square region defined by $-L/2< x_1' < L/2$ and $-L/2< x_2' < L/2$, that is much larger than the illuminated region of the surface. A realization of the surface profile function is then generated numerically, by a two-dimensional extension of the method described in
Refs.~\cite{Maradudin1990,*[{}] [{, Appendix~A.}] Freilikher1997}
on a grid of $(2N+1)^2$ points within this square region of the $x_1'x_2'$ plane~\cite{Simonsen2010-04}. The coordinates of the grid points are
\begin{align}
  \label{eq:53}
  \pvec{x}(\bmath{\ell}) = (\ell_1,\ell_2,0) \Delta x.
\end{align}
In this expression $\ell_1$ and $\ell_2$ are integers that each take the values $-N$, $-N+1$, \ldots, $N-1$, $N$ and which we denote collectively by $\bmath{\ell}$, while $\Delta x=L/(2N+1)$.

We next rewrite Eq.~\eqref{eq:49} as
\begin{align}
  \label{eq:54}
  \NSF(\pvec{x}|\omega)
  &=
    2 \NSF(\pvec{x}|\omega)_{ \mathrm{inc} }
    -
    \frac{1}{2\pi} 
    \sum\limits_{\ell_1'=-N}^{N}
    \sum\limits_{\ell_2'=-N}^{N}
    \;
    \mathcal{P}
    \int\limits_{(\ell_1'-\frac{1}{2})\Delta x}^{(\ell_1'+\frac{1}{2})\Delta x} \dint{x_1'} \!\!\!
    \int\limits_{(\ell_2'-\frac{1}{2})\Delta x}^{(\ell_2'+\frac{1}{2})\Delta x} \dint{x_2'}
    \KernelD(\pvec{x} | \pvec{x}')
    %
    \NSF(\pvec{x}'|\omega),  
\end{align}
where 
\begin{align}
   \KernelD(\pvec{x} | \pvec{x}')
    %
  &=
    \left.
    \left\{
    \exp\left[ \imu \frac{\omega}{c} \big| \pvec{x} - \pvec{x}' \big| \right]
    \left[
    \frac{ \imu \omega/c }{ \big| \pvec{x} - \pvec{x}' \big|^2 }
    -
    \frac{1}{ \big| \pvec{x} - \pvec{x}' \big|^3 }
    \right]
    \left[
    - \left( \pvec{x} - \pvec{x}' \right) \cdot \bmath{\nabla}\zeta(\pvec{x})
    + \left( x_3 - x_3' \right)
    \right]
    \right\}
    \right|_{ \substack{ x_3=\zeta(\pvec{x}) \\ x_3'=\zeta(\pvec{x}')  } }.
\end{align}

We assume that $\NSF(\pvec{x}'|\omega)$ is a slowly varying function  fucntion of $x_1'$ and $x_2'$ in each of the intervals $(\ell_1'-\frac{1}{2})\Delta x < x_1'< (\ell_1'+\frac{1}{2})\Delta x$ and $(\ell_2'-\frac{1}{2})\Delta x < x_2'< (\ell_2'+\frac{1}{2})\Delta x$. We therefore evaluate it at the midpoint of each of these regions, remove it from the integral, and make the change of variable $\pvec{x}'=\pvec{x}(\bmath{\ell}')+\pvec{u}$. The result is the equation
\begin{align}
  \label{eq:56}
   \NSF(\pvec{x}|\omega)
  &=
    2 \NSF(\pvec{x}|\omega)_{ \mathrm{inc} }
    -
    \frac{1}{2\pi}
    \sum\limits_{\ell_1'=-N}^{N}
    \sum\limits_{\ell_2'=-N}^{N}
    \left\{
    \mathcal{P}
    \int\limits_{-\frac{\Delta x}{2}}^{\frac{\Delta x}{2}} \dint{u_1'} \!\!\!
    \int\limits_{-\frac{\Delta x}{2}}^{\frac{\Delta x}{2}} \dint{u_2'}
    \KernelD\big(\pvec{x} | \pvec{x}(\bmath{\ell}')+\pvec{u} \big)
    \right\}
    \NSF\big( \pvec{x}(\bmath{\ell}')|\omega \big).    
\end{align}
To obtain the integral in braces to the lowest order in $\Delta x$, we expand $\KernelD(\pvec{x} | \pvec{x}(\bmath{\ell}')+\pvec{u} )$ in powers of $\pvec{u}$ and keep only the zero-order term. The result is
\begin{align}
  \label{eq:57}
  \NSF(\pvec{x}|\omega)
  &=
    2 \NSF(\pvec{x}|\omega)_{ \mathrm{inc} }
    -
    \frac{1}{2\pi}
    \sum\limits_{\ell_1'=-N}^{N}
    \sum\limits_{\ell_2'=-N}^{N}
    \left( \Delta x \right)^2
    \mathcal{P}
    \KernelD\big(\pvec{x} | \pvec{x}(\bmath{\ell}') \big)
    \NSF\big(\pvec{x}(\bmath{\ell}')|\omega \big).    
\end{align}
We finally set $\pvec{x}=\pvec{x}(\bmath{\ell})$, and obtain the matrix equation satisfied by the $\NSF(\pvec{x}(\bmath{\ell})|\omega)$
\begin{align}
  \label{eq:58}
  \NSF( \pvec{x}(\bmath{\ell}) |\omega)
  &=
    2 \NSF( \pvec{x}(\bmath{\ell}) |\omega)_{ \mathrm{inc} }
    -
    \frac{\left( \Delta x \right)^2}{2\pi}
    \sum\limits_{\ell_1'=-N}^{N}{\hspace{-1.7ex}}^\prime\hspace{1.2ex}
    \sum\limits_{\ell_2'=-N}^{N}{\hspace{-1.7ex}}^\prime\hspace{1.2ex}
    \KernelD\big( \pvec{x}(\bmath{\ell}) | \pvec{x}(\bmath{\ell}') \big)
    \NSF\big(\pvec{x}(\bmath{\ell}')|\omega \big),
    \notag \\
  &
    \qquad \qquad \qquad \qquad \qquad  \qquad \qquad \qquad \qquad \qquad  \ell_{1,2}=-N,-N+1,\ldots,N-1,N.
\end{align}
The primes on the summations indicate that the terms with $\pvec{x}(\bmath{\ell}')=\pvec{x}(\bmath{\ell})$ are omitted. It is in this way that the Cauchy principle value of the integral is evaluated.

\medskip
Turning now to the case of scattering from a Neumann surface, we begin by rewriting Eq.~\eqref{eq:52} as
\begin{align}
  \label{eq:59}
  \SF(\pvec{x}|\omega)
  &=
    2 \SF(\pvec{x}|\omega)_{ \mathrm{inc} }
    +
    \frac{1}{2\pi} 
    \sum\limits_{\ell_1'=-N}^{N}
    \sum\limits_{\ell_2'=-N}^{N}
    \;
    \mathcal{P}
    \int\limits_{(\ell_1'-\frac{1}{2})\Delta x}^{(\ell_1'+\frac{1}{2})\Delta x} \dint{x_1'} \!\!\!
    \int\limits_{(\ell_2'-\frac{1}{2})\Delta x}^{(\ell_2'+\frac{1}{2})\Delta x} \dint{x_2'}
    \KernelN(\pvec{x} | \pvec{x}')
    %
    \SF(\pvec{x}'|\omega),    
\end{align}
where 
\begin{align}
  \label{eq:60}
  \KernelN(\pvec{x} | \pvec{x}')
    %
  &=
    \left.
    \left\{
    -\exp\left[ \imu \frac{\omega}{c} \big| \pvec{x} - \pvec{x}' \big| \right]
    \left[
    \frac{ \imu \omega/c }{ \big| \pvec{x} - \pvec{x}' \big|^2 }
    -
    \frac{1}{ \big| \pvec{x} - \pvec{x}' \big|^3 }
    \right]
    \left[
    - \left( \pvec{x} - \pvec{x}' \right) \cdot \bmath{\nabla}'\zeta(\pvec{x}')
    + \left( x_3 - x_3' \right)
    \right]
    \right\}
    \right|_{ \substack{ x_3=\zeta(\pvec{x}) \\ x_3'=\zeta(\pvec{x}')  } }.
\end{align}
We next assume that $\SF(\pvec{x}'|\omega)$ is a slowly varying function of $x_1'$ and $x_2'$ in each of the intervals $(\ell_1'-\frac{1}{2})\Delta x < x_1'< (\ell_1'+\frac{1}{2})\Delta x$ and $(\ell_2'-\frac{1}{2})\Delta x < x_2'< (\ell_2'+\frac{1}{2})\Delta x$. Then we evaluate it at the midpoint of these regions, remove it from the integral, and make the change of variable $\pvec{x}'=\pvec{x}(\bmath{\ell}')+\pvec{u}$. In this way we obtain the equation
\begin{align}
  \label{eq:61}
   \SF(\pvec{x}|\omega)
  &=
    2 \SF(\pvec{x}|\omega)_{ \mathrm{inc} }
    +
    \frac{1}{2\pi}
    \sum\limits_{\ell_1'=-N}^{N}
    \sum\limits_{\ell_2'=-N}^{N}
    \left\{
    \mathcal{P}
    \int\limits_{-\frac{\Delta x}{2}}^{\frac{\Delta x}{2}} \dint{u_1'} \!\!\!
    \int\limits_{-\frac{\Delta x}{2}}^{\frac{\Delta x}{2}} \dint{u_2'}
    \KernelN\big(\pvec{x} | \pvec{x}(\bmath{\ell}')+\pvec{u} \big)
    \right\}
    \SF\big( \pvec{x}(\bmath{\ell}')|\omega \big).      
\end{align}
We finally set $\pvec{x}=\pvec{x}(\bmath{\ell})$, and evaluate the integral in the braces to the lowest order in $\Delta x$. The result is a matrix equation for $\SF(\pvec{x}(\bmath{\ell})|\omega)$
\begin{align}
  \label{eq:62}
  \SF( \pvec{x}(\bmath{\ell}) |\omega)
  &=
    2 \SF( \pvec{x}(\bmath{\ell}) |\omega)_{ \mathrm{inc} }
    +
    \frac{\left( \Delta x \right)^2}{2\pi}
    \sum\limits_{\ell_1'=-N}^{N}{\hspace{-1.7ex}}^\prime\hspace{1.2ex}
    \sum\limits_{\ell_2'=-N}^{N}{\hspace{-1.7ex}}^\prime\hspace{1.2ex}
    \KernelN\big( \pvec{x}(\bmath{\ell}) | \pvec{x}(\bmath{\ell}') \big)
    \SF\big(\pvec{x}(\bmath{\ell}')|\omega \big),
    \notag \\ 
  &
    \qquad \qquad \qquad \qquad \qquad  \qquad \qquad \qquad \qquad \qquad  \ell_{1,2}=-N,-N+1,\ldots,N-1,N.
\end{align}
Again, primes on the summations indicate that the terms with $\pvec{x}(\bmath{\ell}')=\pvec{x}(\bmath{\ell})$ are omitted.

\section{\label{sec:results}Results and discussion}

Rigorous computer simulations were carried out to obtain the field that is scattered from \textit{isotropic} or \textit{anisotropic} randomly rough Dirichlet or Neumann surfaces. These calculations were performed by numerically solving the inhomogeneous integral equations~\eqref{eq:49} and \eqref{eq:52} by the method outlined in Sec.~\ref{sec:NumSol}; that is, we solved the linear system of equation in Eqs.~\eqref{eq:58} and \eqref{eq:62}. From their solutions, the reflection amplitudes for Dirichlet or Neumann surfaces  were calculated from Eqs.~\eqref{eq:20} and \eqref{eq:23}, respectively, and the results were subsequently used to obtain the mean DRCs defined in Eq.~\eqref{eq:40}. The randomly rough surfaces were assumed to constitute  a Gaussian random process that is characterized by the Gaussian surface height autocorrelation function $W(\pvec{x})$ of the form~\eqref{eq:4}. Realizations of the randomly rough surfaces were generated by the Fourier filtering method as described in Ref.~\onlinecite{Simonsen2010-04}. If nothing is said to indicate otherwise, the edges of the square region of the $x_1x_2$-plane covered by the rough surface were $L=21\lambda$ with $\lambda$ the wavelength of the incident beam, and the half-width of the incident beam, given by the expression in Eq.~\eqref{eq:24}, was $w=L/3=7\lambda$. Moreover, the resulting linear set of equations was solved using the stabilized biconjugated gradient~(BiCGStab) iterative method~\cite{Vorst1992}, with the matrix-vector multiplications that it requires performed using routines from BLAS~(Basic Linear Algebra Subprograms)~\cite{Blackford2002}.

\subsection{Isotropic surfaces }

In the first set of calculations that we performed it was assumed that the rms-roughness of the surface was $\delta=\lambda$ and the correlation length of the isotropic surface was $a=2\lambda$. For an isotropic surface, one takes $a_1=a_2\equiv a$ in Eqs.~\eqref{eq:4} and \eqref{eq:5}. These roughness parameters are identical to those used by Tran and Maradudin~\cite{Tran1992,Tran1992} in their initial study of the scattering of scalar waves from rough impenetrable surfaces. We will start by assuming a rough Dirichlet surface so it is the integral equation~\eqref{eq:49}, and the corresponding linear set of equations~\eqref{eq:58}, we want to solve.

\subsubsection{Isotropic Dirichlet surfaces}

\begin{figure}[tbp!]
  \centering
  \includegraphics[width=0.8\textwidth]{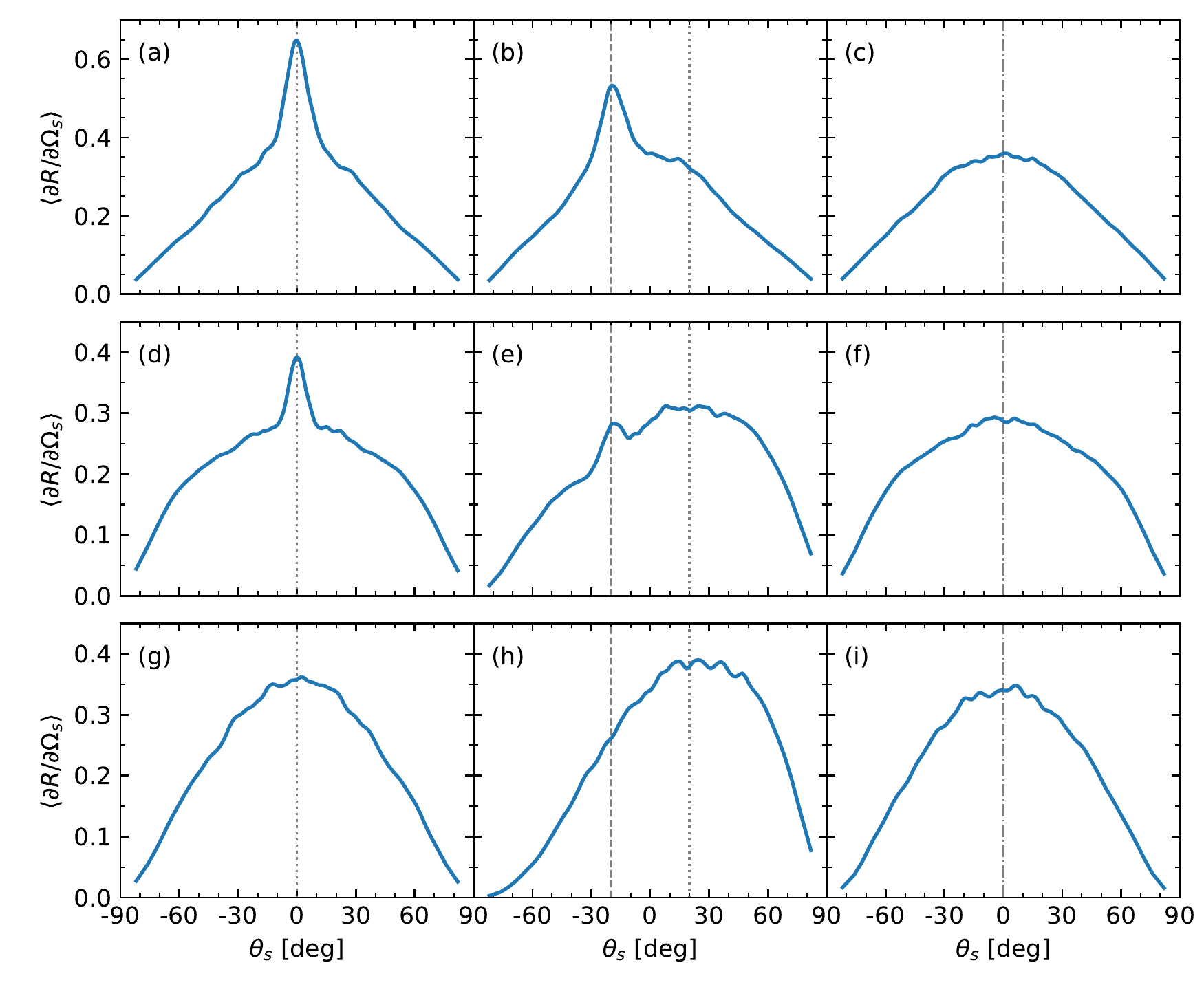}
  \caption{The in-plane and out-of-plane dependencies of the mean DRCs as functions of the scattering angle $\theta_s$ for randomly rough, isotropic, Gaussian correlated  Dirichlet surfaces characterized by the surface roughness $\delta=\lambda$ and correlation length $a=2\lambda$~[Figs.~\ref{Fig:Dirichlet_cuts_rms100}(a--c)], $a=3\lambda$~~[Figs.~\ref{Fig:Dirichlet_cuts_rms100}(d--f)], and $a=4\lambda$~~[Figs.~\ref{Fig:Dirichlet_cuts_rms100}(g--i)] when the surfaces are illuminated by Gaussian beams of wavelength $\lambda$. These results were obtained by assuming a square region of the $x_3=0$ plane, of edges $L=21\lambda$, covered by the randomly rough surface that was discretized using the intervals $\Delta x=\lambda/10$ [$a=2\lambda$] and $\Delta x=\lambda/8$ [$a>2\lambda$]. The half-width of the Gaussian incident beam was $w=L/3=7\lambda$, the polar angle of incidence was $\theta_0=\ang{0}$ or \ang{20}, and the azimuthal angle of incidence was  $\phi_0=\ang{0}$. The different panels correspond either to the in-plane or out-of-plane dependence of the mean DRCs and to different choices for the polar angle of incidence; the in-plane dependence of the mean DRCs are presented in 
Figs.~\ref{Fig:Dirichlet_cuts_rms100}(a,\,d,\,g)  and ~\ref{Fig:Dirichlet_cuts_rms100}(b,\,e,\,h) for $\theta_0=\ang{0}$ and \ang{20}, respectively, while the out-of plane dependence of the mean DRCs for $\theta_0=\ang{20}$ depicted in Figs.~\ref{Fig:Dirichlet_cuts_rms100}(c,\,f,\,i). The results for the mean DRCs that we report were obtained by averaging over an ensemble consisting of at least $N_\zeta=\num{4000}$ realizations of the surface profile function. As a guide to the eye, in each of the in-plane panels we have indicated  the backscattering (retroreflection) and the specular directions by vertical dashed and dotted lines, respectively. Moreover, in out-of-plane panels the directions corresponding to $\theta_s=\ang{0}$ have been marked by vertical dash-dotted lines.} 
  \label{Fig:Dirichlet_cuts_rms100}
\end{figure}

Figure~\ref{Fig:Dirichlet_cuts_rms100}(a) shows the in-plane angular dependence of the mean DRC of the scattered wave when a scalar Gaussian beam is incident normally on the rough surface. This distribution, up to statistical fluctuations, displays a reflection symmetry ($\theta_s\leftrightarrow -\theta_s$) with respect to the angle of scattering $\theta_s=\ang{0}$. Moreover, this distribution is equal to the corresponding out-of-plane distribution, for a normally incident beam, and the latter distribution is therefore not shown. When the beam instead is incident on the rough surface at the polar angle of incidence $\theta_0=\ang{20}$, we obtain the in-plane and out-of-plane mean DRCs depicted in Figs.~\ref{Fig:Dirichlet_cuts_rms100}(b)--(c), respectively. The vertical dashed and dotted lines in Fig.~\ref{Fig:Dirichlet_cuts_rms100} correspond to the 
backscattering and specular directions, respectively, and these lines are included as a guide to the eye. Well defined peaks in the scattered intensity distributions about the retroreflection directions $\theta_s=-\theta_0$ are observed in  Figs.~\ref{Fig:Dirichlet_cuts_rms100}(a)--(b). These peaks are \emph{enhanced backscattering peaks} that result from the constructive interference of volume waves that are scattered multiple times in the grooves of the strongly rough surface~\cite{Tran1992,Mendez1987,Simonsen2004-3}. This wave phenomenon was first observed experimentally in the scattering of light from strongly rough metal  surfaces~\cite{Mendez1987}.  We therefore stress that the peak located at  $\theta_0=\ang{0}$ in Fig.~\ref{Fig:Dirichlet_cuts_rms100}(a), for instance, is not due to coherent (specular) scattering; in fact, the contribution from coherently scattered waves to the angular integral of the mean DRC~[see Eq.~\eqref{eq:energy-conservation} below] is less than $0.09\%$, and this is below the level of precision we have in these simulations for the level of  discretization assumed in performing the calculations. Hence, the mean DRC for this and the other sets of roughness parameters that we will assume are due to waves that are scattered incoherently (diffusely) by the surface roughness. Figure~\ref{Fig:Dirichlet_cuts_rms100}(c) presents the corresponding out-of-plane dependence [$\phi_s=\phi_0\pm\ang{90}$] of the mean DRC for the polar angle of incidence $\theta_0=\ang{20}$. The maximum scattered intensity is found around $\theta_s=\ang{0}$, indicated by the vertical dash-dotted line in Fig.~\ref{Fig:Dirichlet_cuts_rms100}(c). From this figure we observe that as $|\theta_s|$ increases, the scattered intensity drops off from its maximum value at $\theta_s=\ang{0}$ and the distribution is approximately reflection symmetric with respect to the normal scattering direction; the scattered intensity distribution from an isotropic surface is expected to display such left-right symmetry with respect to the plane of incidence.

The next set of roughness parameters we consider consists of $\delta=\lambda$ (unchanged) and $a=3\lambda$. The resulting in-plane and out-of-plane dependencies of the mean DRCs are presented in Figs.~\ref{Fig:Dirichlet_cuts_rms100}(d)--(e) and \ref{Fig:Dirichlet_cuts_rms100}(f), respectively, for the polar angles of incidence $\theta_s=\ang{0}$ and \ang{20}. Qualitatively these results show several features that are similar to what is observed when the correlation length is $a=2\lambda$~[Fig.~\ref{Fig:Dirichlet_cuts_rms100}(a)--(c)]. For instance, for both of the polar angles of incidence $\theta_0=\ang{0}$ and \ang{20}, enhanced backscattering peaks are observed at $\theta_s=-\theta_0$. However, there are also differences between the mean DRCs obtained for the two sets of roughness parameters. For instance, Fig.~\ref{Fig:Dirichlet_cuts_rms100}(e) shows that the in-plane scattered intensity has a maximum in the forward scattering plane ($\theta_s>\ang{0}$); this is not the case when  $a=2\lambda$ [Fig.~\ref{Fig:Dirichlet_cuts_rms100}(b)] for which the maximum of the in-plane scattered intensity is found in the backscattering direction and therefore in the backscattering plane ($\theta_s<\ang{0}$). For normal incidence, a comparison of the in-plane mean DRCs in Figs.~\ref{Fig:Dirichlet_cuts_rms100}(a) and \ref{Fig:Dirichlet_cuts_rms100}(d) reveals that the backscattered intensity is higher when $a=2\lambda$ than what it is when $a=3\lambda$. A more detailed study of the simulation results in the same figures show that the ratio of the scattered intensity in the retroreflection direction $\theta_s=-\theta_0$ to the intensity of the background at its position is higher when $a=2\lambda$ then what it is when $a=3\lambda$. If single-scattering contributions to the scattered intensity can be neglected around the retroreflection direction, the intensity of the backscattering peak is expected to be twice that of the background intensity~\cite{Simonsen2004-3,Maradudin1990}; this implies that the intensity ratio we defined above should have the value two. From the results in  Figs.~\ref{Fig:Dirichlet_cuts_rms100}(a)--(b)~[$a=2\lambda$] and \ref{Fig:Dirichlet_cuts_rms100}(d)--(e)~[$a=3\lambda$]  it is observed that the intensity ratio is a little smaller than two when $a=2\lambda$ and smaller than this value when $a=3\lambda$. These findings we take as an indication that multiple scattering processes contribute more significantly to the scattered field for the former set of roughness parameters than for the latter set of roughness parameters. We find also, for both $\theta_0=\ang{0}$ and $\ang{20}$, that the enhanced backscattering peak is wider in Figs.~\ref{Fig:Dirichlet_cuts_rms100}(a)--(b) than in Figs.~\ref{Fig:Dirichlet_cuts_rms100}(d)--(e), that is, for the shorter transverse correlation length of the surface roughness. This is consistent with what has been observed previously for the scattering of light from one-dimensional randomly rough surfaces~\cite{Maradudin1990}. Furthermore, the results in Figs.~\ref{Fig:Dirichlet_cuts_rms100}(a) and \ref{Fig:Dirichlet_cuts_rms100}(d)  show that  the entire in-plane scattered intensity distribution is broader for the  case when the correlation length is $a=3\lambda$ than for the case when it is $a=2\lambda$.

Finally, Figs.~\ref{Fig:Dirichlet_cuts_rms100}(g)--(i) present computer simulation results for the in-plane or out-of-plane mean DRCs obtained when the correlation length of the rough Dirichlet surface is $a=4\lambda$ with the remaining roughness and numerical parameters having unchanged values; this is twice the value of the correlation length assumed in obtaining the results presented in, for instance, Figs.~\ref{Fig:Dirichlet_cuts_rms100}(a)--(c). Contrary to what was found when $a=2\lambda$ and $a=3\lambda$ (with $\delta=\lambda$), the results in Figs.~\ref{Fig:Dirichlet_cuts_rms100}(g)--(i) for $a=4\lambda$ show no well-defined enhanced backscattering peaks in the scattered intensity distributions; in this case, the local slopes~\cite{Simonsen2004-3} of the Gaussian surface are simply too small to allow for any significant contribution from multiple scattering of volume waves. Instead the in-plane and out-of-plane scattered intensity distributions are found to be featureless around the backscattering and specular directions. Moreover, when $\theta_0=\ang{20}$ the maximum scattered intensity is located in the forward scattering plane ($\theta_s>\ang{0}$), see Fig.~\ref{Fig:Dirichlet_cuts_rms100}(h).

\begin{figure}[tbph!]
  \centering
  \includegraphics[width=0.8\textwidth]{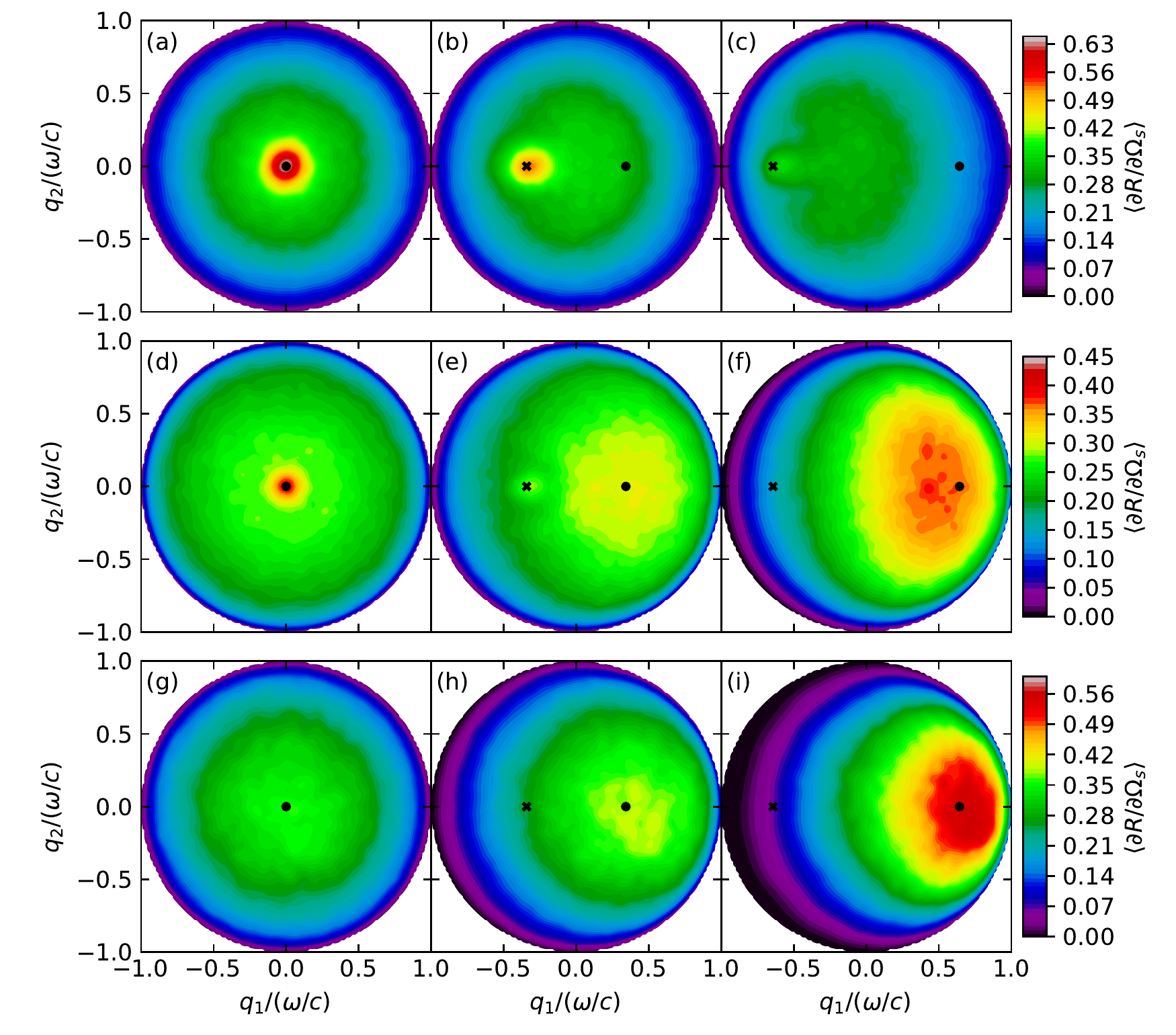}
\caption{The full angular dependence of the mean DRCs for randomly rough, isotropic, Gaussian correlated  Dirichlet surfaces as functions of the in-plane scattered wave vector $\pvec{q}$.
  In all cases the azimuthal angle of incidence was $\phi_0=\ang{0}$ and the polar angles of incidence were $\theta_0=\ang{0}$~[Figs.~\ref{Fig:Dirichlet_rms100}(a,\,d,\,g)]; \ang{20}~[Figs.~\ref{Fig:Dirichlet_rms100}(b,\,e,\,h)]; and \ang{40}~[Figs.~\ref{Fig:Dirichlet_rms100}(c,\,f,\,i)].
  The roughness parameters were those of Fig.~\ref{Fig:Dirichlet_cuts_rms100}, that is, for all cases the roughness of the surface was   $\delta=\lambda$, and the correlation length was $a=2\lambda$~[Figs.~\ref{Fig:Dirichlet_rms100}(a--c)]; 
  $a=3\lambda$~[Figs.~\ref{Fig:Dirichlet_rms100}(d--f)]; and
  $a=4\lambda$~[Figs.~\ref{Fig:Dirichlet_rms100}(g--i)].
  The remaining scattering and numerical parameters are identical to those presented in the caption of Fig.~\ref{Fig:Dirichlet_cuts_rms100}. As a guide to the eye, the positions of the backscattering and the specular directions have been indicated by black crosses and filled circles, respectively.}
  \label{Fig:Dirichlet_rms100}
\end{figure}

Based on the angular dependence of the  in-plane mean DRCs in Fig.~\ref{Fig:Dirichlet_cuts_rms100} alone, one can strictly speaking not attribute the features observed at $\theta_s=-\theta_0$ in these distributions to the backscattering phenomenon. To reach this conclusion, one is required  to  demonstrate that these features are actually peaks in the retroreflection direction and not, for instance, ridges as were recently observed in the mean DRC for cross-polarized light scattered from a rough  perfect electric conducting surface~\cite{Simonsen2009-1}. The full angular distribution of the mean DRCs for the scattering of scalar waves are presented in Fig.~\ref{Fig:Dirichlet_rms100} for polar angles of incidence $\theta_0=\ang{0}$, \ang{20}, \ang{40}, and for the three sets of roughness parameters for the Dirichlet surfaces used to produce the results in Fig.~\ref{Fig:Dirichlet_cuts_rms100}. Figure~\ref{Fig:Dirichlet_rms100} shows several examples of well-defined peaks in the mean DRCs about the retroreflection direction [$\pvec{q}=-\pvec{k}$]; in particular, these results demonstrate explicitly that the peaks at $\theta_s=-\theta_0$ in Figs.~\ref{Fig:Dirichlet_cuts_rms100}(a)--(b) and Figs.~\ref{Fig:Dirichlet_cuts_rms100}(d)--(e) are enhanced backscattering peaks. Moreover, the comments that were made about the results in Fig.~\ref{Fig:Dirichlet_cuts_rms100} regarding the width of the angular distributions are readily seen to apply to the results presented in Fig.~\ref{Fig:Dirichlet_rms100}; indeed the distributions in Figs.~\ref{Fig:Dirichlet_rms100}(a)--(b) that correspond to the correlation length $a=2\lambda$ are \emph{less} wide than the angular distributions in Figs.~\ref{Fig:Dirichlet_rms100}(a)--(b) [$a=3\lambda$]. However, the most interesting observation to be made from the results in  Fig.~\ref{Fig:Dirichlet_rms100} is how the backscattering peaks depend on the polar angle of incidence $\theta_0$ and the correlation length of the surface when the surface roughness is the same. The results presented in Figs.~\ref{Fig:Dirichlet_rms100}(a)--(c) correspond to $a=2\lambda$ and display well defined enhanced backscattering peaks for all the polar angles of incidence $\theta_0=\ang{0}$, \ang{20}, and \ang{40}. On the other hand, when the correlation length of the surface is increased from $a=2\lambda$ to $a=3\lambda$, the mean DRCs in Figs.~\ref{Fig:Dirichlet_rms100}(d)--(e), corresponding to the polar angles of incidence $\theta_0=\ang{0}$, \ang{20}, and \ang{40}, respectively, show enhanced backscattering peaks that gradually disappear with increasing polar angles of incidence. For instance, in Fig.~\ref{Fig:Dirichlet_rms100}(f)~[$\theta_0=\ang{40}$] no backscattering peak is observed, while such a peak is observed in the mean DRC for the same polar angle of incidence when $a=2\lambda$~[Fig.~\ref{Fig:Dirichlet_rms100}(c)].

\smallskip
At this stage it should be commented that from the results of the full angular dependence of the mean DRC in Fig.~\ref{Fig:Dirichlet_rms100} one can check the satisfaction of the energy conservation; this is often referred to as \emph{unitarity}.  Since the Dirichlet and Neumann surfaces are impenetrable to scalar waves, all energy incident on them has to be reflected away from them. From the definition of the DRCs in Eqs.~\eqref{eq:39} and \eqref{eq:40} it follows that 
\begin{align}
  {\mathcal U}(\pvec{k}) =
  \int\limits_{q_\parallel<\omega/c}
  d\Omega_s \,
  \left<
  \frac{
  \partial R(\pvec{q}|\pvec{k})
  }{
  \partial \Omega_s
  }
  \right>
  = 1.
  \label{eq:energy-conservation}
\end{align}
The relation in Eq.~\eqref{eq:energy-conservation} is a consequence of  energy conservation. Strictly speaking energy conservation should be satisfied individually for each of the DRCs that enters into the calculation of the mean DRC, but this possibility we will not explore here. It should be noted, that relation \eqref{eq:energy-conservation} is a necessary but not a sufficient condition for correct results. However, we have found that the satisfaction of Eq.~\eqref{eq:energy-conservation} is a good criterion for gauging the quality of simulation results and, for instance, if the discretization interval used in performing the calculations is small enough for the assumed roughness parameters. For the results for normal incidence presented in Fig.~\ref{Fig:Dirichlet_rms100}, and therefore also in Fig.~\ref{Fig:Dirichlet_cuts_rms100}, we found that energy conservation was satisfied to within an error of no more than \num{2E-3} for the numerical parameters [see caption in Fig.~\ref{Fig:Dirichlet_rms100}] assumed in performing these calculations. For non-normal incidence, the error in the satisfaction of the energy conservation condition was somewhat larger, partly due to the increased footprint of the incident beam on the mean surface; for all the calculations reported in Fig.~\ref{Fig:Dirichlet_rms100} this error was always smaller than \num{E-2} which testifies to the quality of the simulation results that we have obtained.      

\medskip
At this stage we should mention the computational resources needed to produce the simulation results presented in Fig.~\ref{Fig:Dirichlet_rms100}. For instance, the results in Fig.~\ref{Fig:Dirichlet_rms100}(a) took \SI{598}{s}, or almost \SI{10}{min}, of computer time to produce per surface realization when the simulations were performed on a single core of an Intel® i7-960 processor (8M Cache, \SI{3.20}{GHz}). This time was spent in the following manner: \SI{385}{s} for setting up the linear system of equations; \SI{125}{s} to solve it using the BiCGStab iterative solver (for one angle of incidence); \SI{44}{s} to calculate the reflection amplitudes in Eqs.~\eqref{eq:20} and \eqref{eq:23}; and finally another \SI{44}{s} to obtain the DRC and mean DRC defined by Eqs.~\eqref{eq:39} and \eqref{eq:40}. For the simulations that we performed, the code was OpenMP parallelized, in which case the wall time for the same simulation was reduced; using the processor given above and running on all 8~cores resulted in a wall time of \SI{5.3}{min} for performing the same calculations. The memory footprint of the simulations was almost \num{15}{Gb}, of which the majority went to storing the dense complex matrix for the \num{44100} linear equations.


\subsubsection{Isotropic Neumann surfaces}
\begin{figure}[tbp!]
  \centering
  \includegraphics[width=0.8\textwidth]{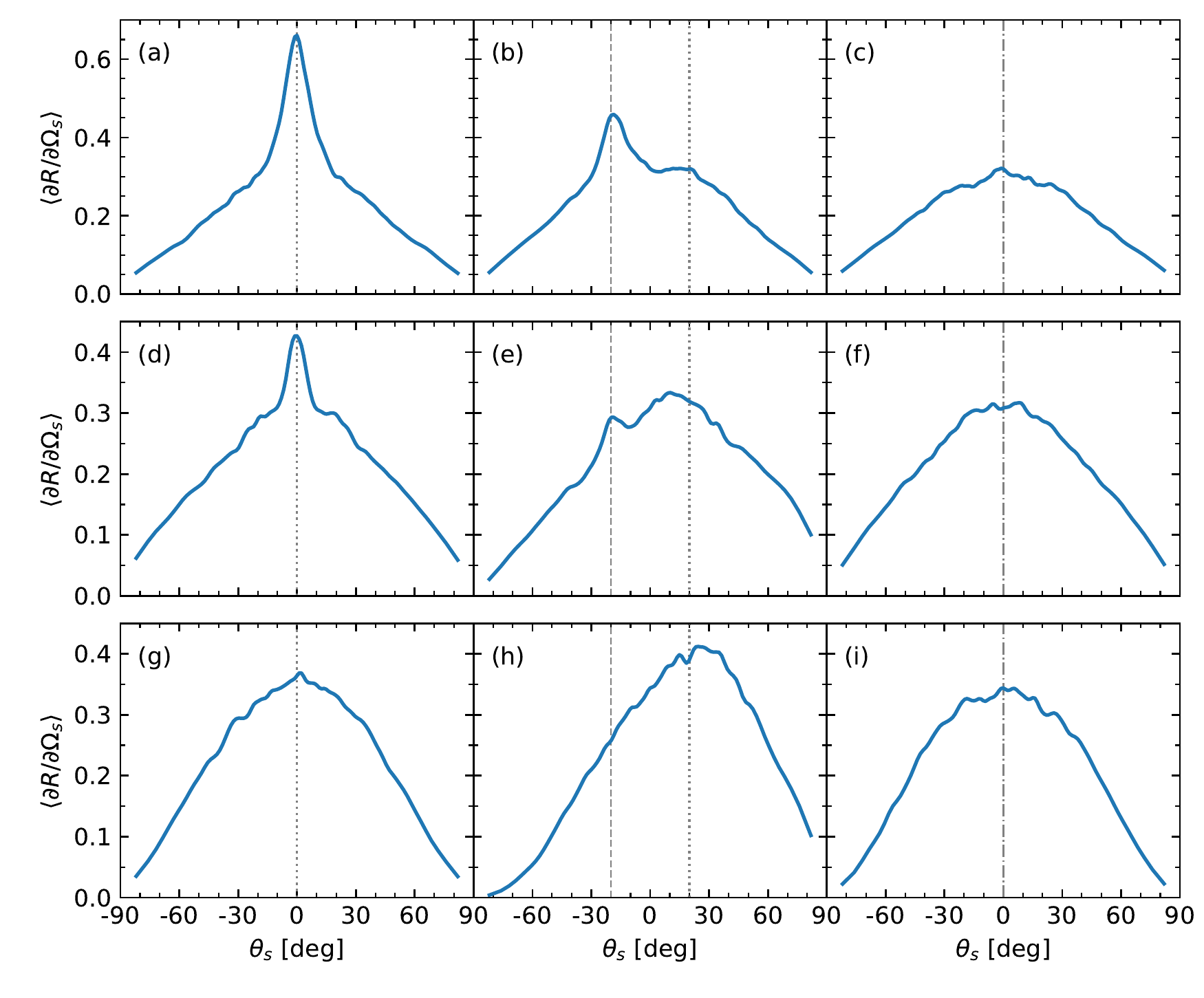}
  \caption{Same as Fig.~\protect\ref{Fig:Dirichlet_cuts_rms100} but for Neumann surfaces.}
  \label{Fig:Neumann_cuts_rms100}
\end{figure}

\begin{figure}[tbp!]
  \centering
\includegraphics[width=0.8\textwidth]{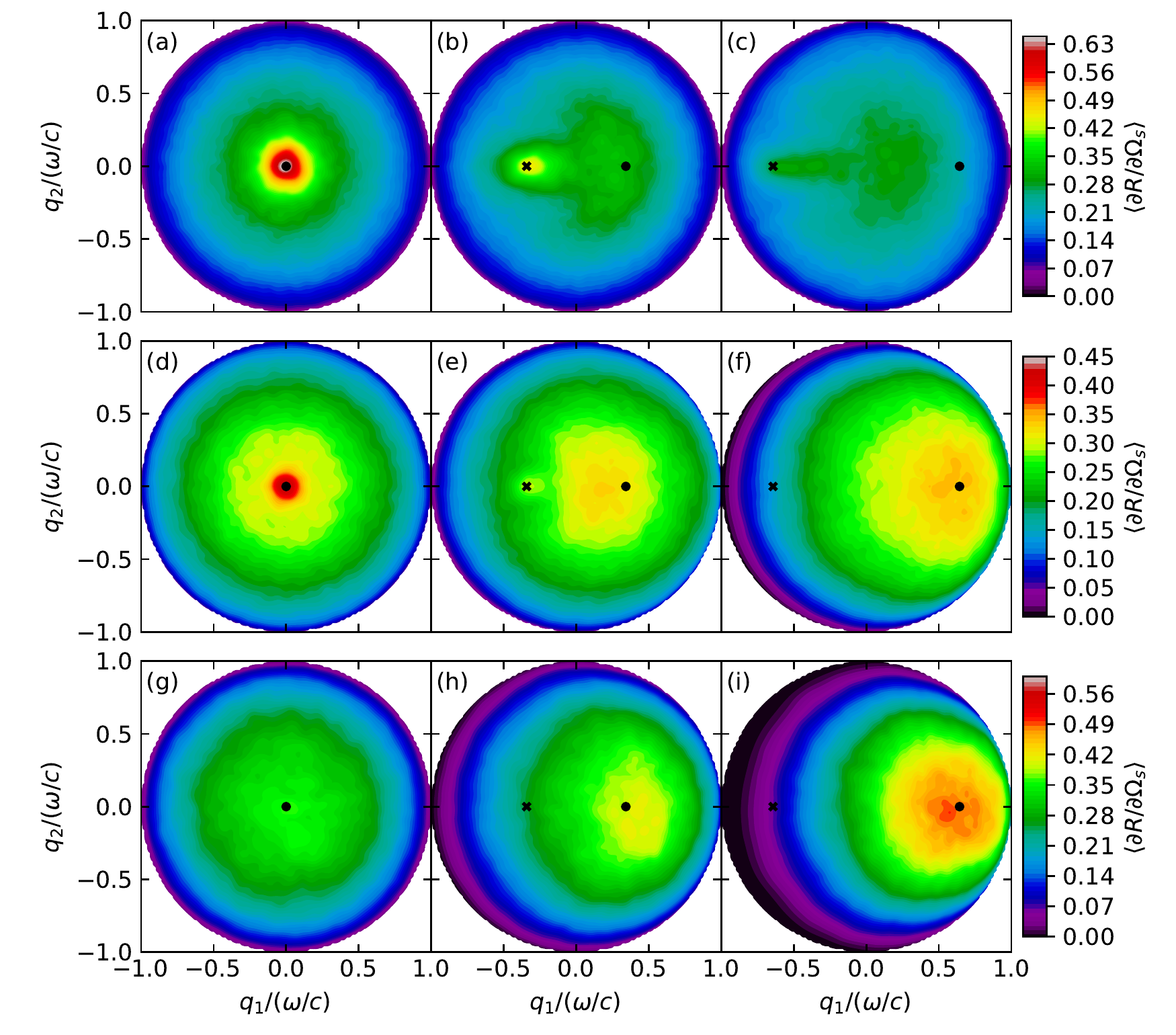}
\caption{Same as Fig.~\protect\ref{Fig:Dirichlet_rms100} but for Neumann surfaces.}
  \label{Fig:Neumann_rms100}
\end{figure}

We now address rough Neumann surfaces and the scattering of a scalar beam from them. The parameters of the isotropic randomly rough surfaces we will take to be  identical to those assumed for the Dirichlet case~[Figs.~\ref{Fig:Dirichlet_cuts_rms100} and \ref{Fig:Dirichlet_rms100}]. The in-plane and out-of-plane angular dependencies of the mean DRCs for rough Neumann surfaces,
obtained on the basis of solving  the linear system Eq.~\eqref{eq:62}, are presented in Fig.~\ref{Fig:Neumann_cuts_rms100}. The corresponding full angular dependencies of the mean DRCs are depicted in  Fig.~\ref{Fig:Neumann_rms100} where results for $\theta_0=\ang{40}$ have also been included. At least for the roughness parameters and polar angles of incidence that we have assumed here, the mean DRCs for both Neumann and Dirichlet surfaces are qualitatively rather similar. For instance, an enhanced backscattering peak is observed around $\pvec{q}=-\pvec{k}$ in the mean DRC for a Neumann surface if it also is observed in the mean DRC for a  Dirichlet surface with the same roughness parameters. Moreover, the behavior we reported above for the mean DRCs of Dirichlet surfaces for increasing values of $a$ and $\theta_0$, we also find in the case of scattering from Neumann surfaces. The only minor differences between the mean DRCs for Neumann and  Dirichlet surfaces we find worthy of a comment is the difference in shape of the in-plane and out-of-plane distributions for these two kinds of surfaces. In particular, a detailed comparison of the results in Figs.~\ref{Fig:Dirichlet_cuts_rms100}(d)--(f) and Figs.~\ref{Fig:Neumann_cuts_rms100}(d)--(f), corresponding to the scattering from a Dirichlet and Neumann surface of the same surface statistics, reveals that the tails and amplitudes of the distributions are somewhat different. The distributions in the former case are semi-circular, while in the latter case the distributions are more triangular.

\subsubsection{Reflectivity}
\begin{figure}[tbph!]
  \centering
  \includegraphics[width=0.5\textwidth]{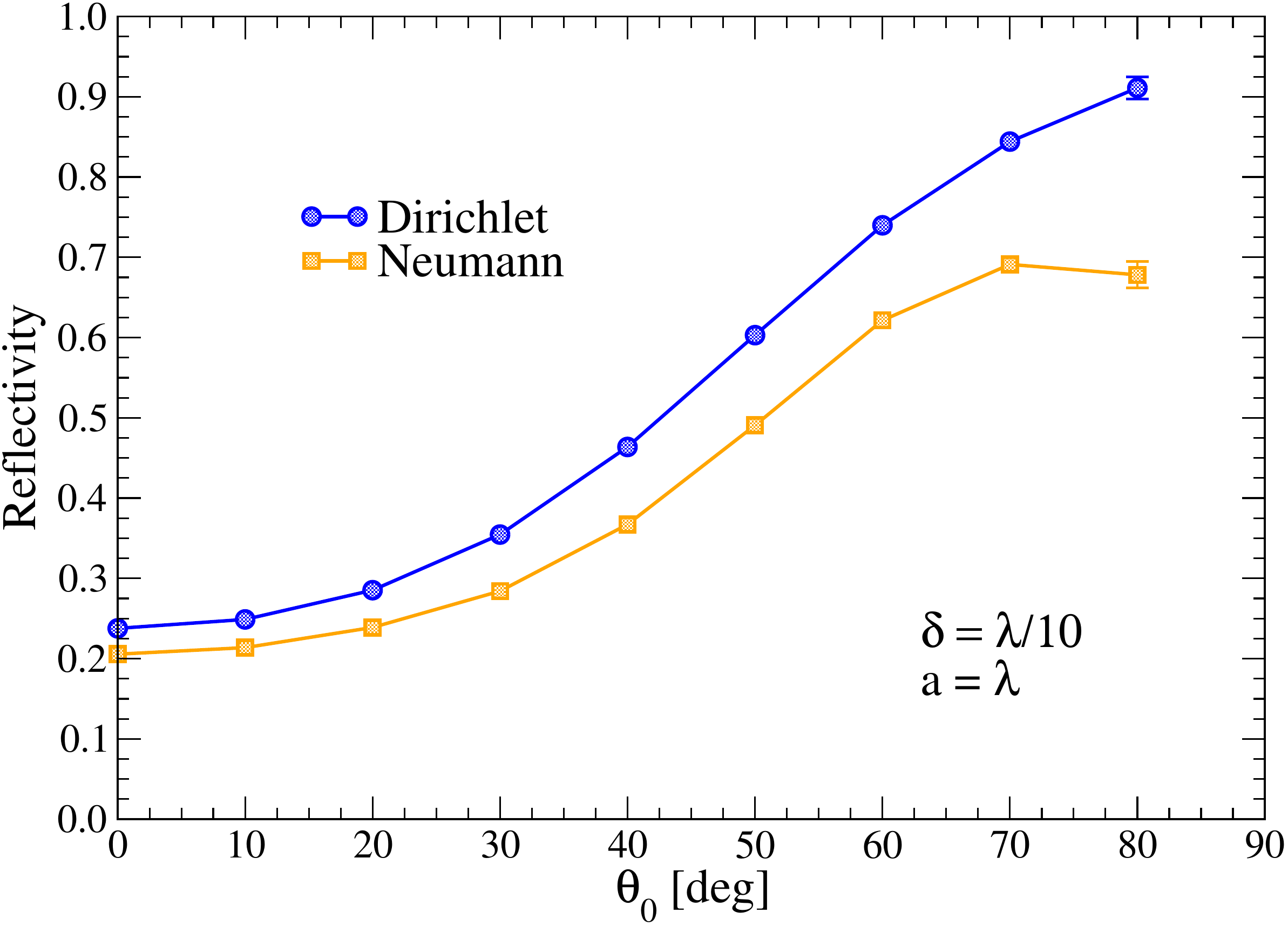}
  \caption{The reflectivity as a function of the polar angle of incidence calculated on the basis of Eq.~\eqref{eq:45} for isotropic Gaussian correlated Dirichlet and Neumann surfaces of roughness $\delta=\lambda/10$ and correlation length $a=\lambda$. Here $\lambda$ denotes the wavelength of a Gaussian beam of width $w=L/3$ incident on a rough surface of edges $L=32\lambda$. The remaining numerical parameters are identical to those used in obtaining the results of Fig.~\ref{Fig:Dirichlet_rms100}. The error bars, only clearly visible for $\theta_0=\ang{80}$, indicate the absolute deviation from unity of the unitarity obtained in the numerical simulations. The reported results were obtain by averaging the results of \num{50} surface realizations, which was sufficient to achieve convergent results.}
    \label{Fig:Reflectivity}
\end{figure}

For the isotropic surfaces that we have studied until now, the roughness parameters were such that the mean DRCs for both the rough Dirichlet and Neumann surfaces  were fully diffuse. In view of Eq.~\eqref{eq:45} this means that the reflectivity of such surfaces is neglectable and smaller than the error found in the satisfaction of the energy conservation condition~\eqref{eq:energy-conservation}. In order to have a non-vanishing reflectivity, we will now consider less rough surfaces;  in particular, we work with the roughness parameters $\delta=\lambda/10$ and $a=\lambda$, still assuming an isotropic  Gaussian correlation function. For such roughness parameters the mean DRC was calculated for a sequence of lateral wave vector $\pvec{k}=k_\parallel \pvecUnit{k}$ of the incident Gaussian beam where $k_\parallel = (\omega/c) \sin\theta_0$. In particular, the calculation were performed for polar angles of incidence from $\theta_0=\ang{0}$ to \ang{80} in steps of \ang{10}. For each polar angle of incidence, the reflectivity was calculated from Eq.~\eqref{eq:45} by using the coherent component of the mean DRC. The mean DRCs were obtained on the basis of sufficiently many surface realizations so that the calculated reflectivity had converged to at least four decimal places; for the values of the roughness and numerical parameters assumed, \num{75} surface realizations were sufficient to obtain such a convergence. In this way, we obtain the reflectivity curves presented in Fig.~\ref{Fig:Reflectivity}, where the error bars represent $|1-{\mathcal U}(\pvec{k})|$.  From the results presented in this figure one  observes that the reflectivity of the Dirichlet surface is always higher than the reflectivity of the Neumann surface with the same roughness parameters, and this is even the case for normal incidence. Moreover, increasing the polar angle of incidence seems to increase the reflectivity; only for the largest angle of incidence that we consider, $\theta_0=\ang{80}$, there may be an exception to this rule in the Neumann case.  For such large polar angles of incidence the precision in the simulations (see error bars) is simply not good enough to reach a definite conclusion on this issue. In any case, we remark that when the polar angle of incidence is approaching $\ang{90}$ the reflectiveties of both the Dirichlet and the Neumann surfaces should approach unity. Based on the results presented in Fig.~\ref{Fig:Reflectivity}, it is tempting to speculate that the reflectivity of the Dirichlet surface will smoothly approach unity, while the reflectivity of the Neumann surface will go through a local minimum before again increasing and reaching unity for grazing angles of incidence. Such behavior of the reflectivity was recently observed in the scattering of s- and p-polarized light from randomly rough perfectly conducting surfaces~\cite{Simonsen2012-05}. The determination of whether  our speculations are well founded we will leave for a dedicated study that probably will benefit from the use of a plane incident wave.

\subsubsection{Comparison to previous results}
Before leaving the scattering from isotropic surfaces, it should be remarked that
more than 20 years ago, Tran and Maradudin published the initial rigorous computational results for the in-plane mean DRC obtained when normally incident scalar beams are scattered from strongly rough Dirichlet and Neumann surfaces~\cite{Tran1992,Tran1993}. Their calculations were based on the integral equations Eqs.~\eqref{eq:49} and \eqref{eq:52}, which were transformed into a linear set of equations when each realization of the surface profile function $\zeta(\pvec{x})$ of edges $L=16\lambda$ was discretized onto a square grid of $N=64$ points per side ($\Delta x=\lambda/4$), and the resulting linear system of ($N^2=\num{4096}$) equations was solved iteratively by a method based on the Liouville-Neumann series~\cite{Book:Mathews1970}. In the calculation results reported in Refs.~\cite{Tran1992,Tran1993} an isotropic Gaussian correlation function was used, while the values of the roughness parameters, $\delta=\lambda$ and $a=2\lambda$, assumed were identical to those assumed in performing the calculations whose results are presented as the first row of sub-figures in Figs.~\ref{Fig:Dirichlet_cuts_rms100}--\ref{Fig:Neumann_rms100}. The main differences between the results that we report and those reported in Refs.~\cite{Tran1992,Tran1993} are that we use a larger surface,  a wider incident beam, and a smaller discretization interval in performing the calculations, relative to the corresponding parameters used in the initial calculations. Moreover, the linear equation system is solved in a rather different manner. A direct comparison of the results from Figs.~\ref{Fig:Dirichlet_cuts_rms100}(a) and \ref{Fig:Neumann_cuts_rms100}(a) to the results of Figs.~2(b) and 3(b) in Ref.~\onlinecite{Tran1993} reveals a fair agreement between the former and latter sets of results. For instance, all simulation results for $\theta_0=\ang{0}$ predict the amplitude of the mean DRC at $\theta_s=\ang{0}$ to be somewhat above \num{0.6}. However, the results in Ref.~\onlinecite{Tran1993} seem to predict a less broad angular intensity distribution than what we find in the calculations reported here. The observed difference we suspect is caused by the significantly smaller discretization interval used in performing the calculations reported in Ref.~\onlinecite{Tran1993}. With the use of the numerical parameters assumed in this publication, we were able to reproduce rather well the results for a Dirichlet surface reported by Tran and Maradudin~\cite{Tran1993}.

\smallskip
It ought to be remarked that energy conservation, and thus the quality check on the simulation results that follows from it, could not have been performed on the simulation results produced by Tran and Maradudin~\cite{Tran1992,Tran1993} or by Macaskill and Kachoyan~\cite{Macaskill1993}. In none of these publications was the full angular distribution of the scattered intensity calculated, which is required to obtain ${\mathcal U}(\pvec{k})$ defined in Eq.~\eqref{eq:energy-conservation}. Finally, we remark that in the calculations that we performed for $\theta_0=\ang{0}$ assuming the numerical parameters (with $\Delta x=\lambda/4$) from Ref.~\cite{Tran1993}, energy conservation was satisfied within an error of $6.7\%$; for comparison, we recall from the preceding discussion,  that the corresponding results using our numerical parameters (with $\Delta x=\lambda/10$) resulted in an error in the satisfaction of the energy conservation of no more than $0.2\%$.

\subsection{Anisotropic surfaces}

\begin{figure*}[bt!]
  \centering
  \includegraphics[width=\textwidth]{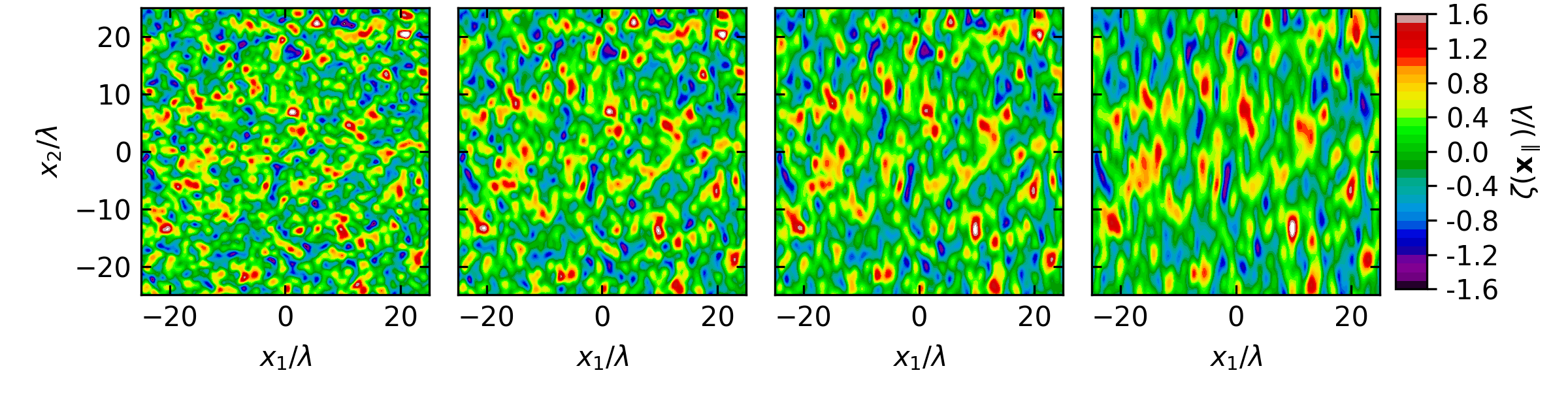}
  \caption{Contour plots of the Gaussian correlated randomly rough surfaces $\zeta(\pvec{x})$  defined by the roughness parameters  $\delta=\lambda/2$, $a_1=\lambda$ and (a)~$a_2=\lambda$ (isotropic surface); (b)~$a_2=\num{1.5}\lambda$; (c)~$a_2=2\lambda$ and (d)~$a_2=3\lambda$. The surface realizations were generated by the Fourier filtering method~\cite{Simonsen2010-04}.  The generation of the surface realizations assumed the \textit{same} uncorrelated random numbers so that one can follow how surface features are affected by the increasing level of anisotropy. } 
  \label{Fig:AnisotropicSurfaces-SurfaceRoughness}
\end{figure*}

\begin{figure*}[tbp!]
  \centering
  \includegraphics[width=0.8\textwidth]{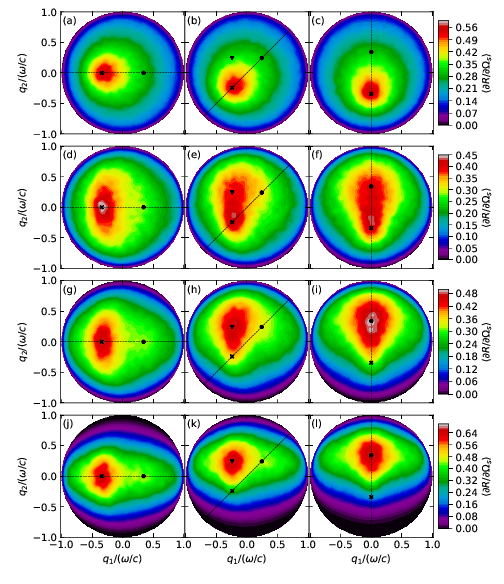}
\caption{The full angular dependence of the mean DRCs for randomly rough, \emph{anisotropic}, Gaussian correlated  Dirichlet surfaces as functions of the in-plane scattered wave vector $\pvec{q}$. The surfaces were illuminated by  Gaussian incident beams of wavelength $\lambda$ and angles of incidence  $(\theta_0,\phi_0)$ were $\theta_0=\ang{20}$ and $\phi_0=\ang{0}$~[1st column; Figs.~\ref{Fig:AnisotropicSurfaces}(a,\,d,\,g,\,j)],
  \ang{45}~[2nd column; Figs.~\ref{Fig:AnisotropicSurfaces}(b,\,e,\,h,\,k)],
and \ang{90}~[3rd column; Figs.~\ref{Fig:AnisotropicSurfaces}(c,\,f,\,i,\,l)].
The surface roughness is $\delta=\lambda/2$  and the surface height autocorrelation function is defined by Eq.~\eqref{eq:4} with correlation lengths $a_1=\lambda$ and $a_2=\lambda$~[1st row]; $a_2=\num{1.5}\lambda$~[2nd row]; $a_2=2\lambda$~[3rd row]; and finally $a_2=3\lambda$~[4th row]. The sampling interval used in performing the numerical simulations was $\Delta x=\lambda/8$. All remaining numerical parameters and the  property of the incident beams are identical
to those assumed in obtaining the results in Fig.~\ref{Fig:Dirichlet_rms100}. The reported results were obtained by averaging over an ensemble consisting of $N_\zeta=\num{4000}$ of surface realizations.
The thin dashed black lines that appear in each of the panels indicate the direction of the plane of incidence. The backscattering and specular directions are marked by black crosses and filled circles. If the projection of the wave vector of incidence is $\pvec{k}=(k_1,k_2,0)$ then the directions $\pvec{q}=(-k_1,k_2,0)$ are indicated by black filled triangles in the 2nd column.} 
  \label{Fig:AnisotropicSurfaces}
\end{figure*}

Until now we have exclusively been dealing with isotropic surfaces. However, many naturally occurring or man made surfaces are anisotropic. Therefore, we now turn our attention to the scattering from anisotropic surfaces. In this case the correlation function $W(\pvec{x})$ that we assume has the form~\eqref{eq:4} with $a_1\neq a_2$. We remark that this form of anisotropy is not the most general one; for instance, the main axes of the anisotropy do not have to be orthogonal to each other, but such a more general case will not be addressed here.

In the computer simulations that we performed, the polar angle of incidence was $\theta_0=\ang{20}$, the surface roughness was $\delta=\lambda/2$, and the correlation length along the $x_1$-direction had the constant value $a_1=\lambda$. The correlation length along the $x_2$-direction was then varied so that $a_2\in\{\lambda,1.5\lambda,2\lambda,3\lambda\}$. A realization of the rough surface for each of these roughness parameters is presented in Fig.~\ref{Fig:AnisotropicSurfaces-SurfaceRoughness}. To better be able to follow how surface features are affected by increasing the level of anisotropy, each of these surface realizations was generated on the basis of the \emph{same} uncorrelated random numbers by the method described in Ref.~\onlinecite{Simonsen2010-04}. We learn from the surface topographies depicted in Fig.~\ref{Fig:AnisotropicSurfaces-SurfaceRoughness} that when $a_2$ is several times larger than $a_1$, the resulting surface topography starts taking the form of a ``randomly corrugated iron'' structure where the grooves of the structure are along the $x_2$-direction.   In the limit that $a_2/\lambda\rightarrow\infty$, or more precisely that $a_2\gg a_1$ with $a_2\gg\lambda$, the surface starts approaching a one-dimensional surface for which the surface profile function $\zeta(\pvec{x})$ will be independent of the spatial coordinate $x_2$. Moreover, the numerical parameters assumed in the simulations and the parameters characterizing the incident beam were identical to those used previously in the study of the scattering from isotropic surfaces~[see captions of Figs.~\ref{Fig:Dirichlet_cuts_rms100} and \ref{Fig:Dirichlet_rms100}]. The value of the azimuthal angle of incidence, $\phi_0$,  we assumed to be \ang{0}, \ang{45}, and \ang{90}. Notice that when we are dealing with the mean DRC there is no need to consider values of $\phi_0$ outside the interval \ang{0} to \ang{90} since the mean DRC for $\phi_0\notin [\ang{0},\ang{90}]$ can be related by symmetry to the mean DRC for a value of $\phi_0\in [\ang{0},\ang{90}]$. However, note that for the DRC that a single surface realization gives rise to, this is no longer true.
    
For later comparison, we start by presenting the full angular dependence of the mean DRC for the isotropic case [$a_1=a_2=\lambda$] for three values of the  azimuthal angle of incidence, $\phi_0=\ang{0}$, \ang{45} and \ang{90}~[Figs.~\ref{Fig:AnisotropicSurfaces}(a)--(c)]. In each panel of Fig.~\ref{Fig:AnisotropicSurfaces} the direction of the plane of incidence has been indicated by a thin black dashed line. Furthermore,  in the same figure, the black crosses and filled black circles represent the backscattering and specular directions, respectively. The results presented in Figs.~\ref{Fig:AnisotropicSurfaces}(a)--(c) all display well-defined enhanced backscattering peaks, and these results are related to each other by azimuthal rotations (up to fluctuations which are caused by the use of a finite number of surface realizations); for instance, the mean DRC corresponding to $\phi_0=\ang{45}$ can be obtained from  the mean DRC for $\phi_0=\ang{0}$~(\ang{90}) by a counter-clockwise rotation through an azimuthal angle of \ang{45}~($-\ang{45}$) about the $q_3$-axis. This is a consequence of the isotropy of the surface and that the polar angle of incidence is the same for the results in Figs.~\ref{Fig:AnisotropicSurfaces}(a)--(c). It should be mentioned that for the roughness parameters we assume the scattering was fully incoherent (diffusive); in fact, the reflectivity was of the order of \num{E-4}, which is on a par with the error level in these simulations.

%
%
We now turn to Gaussian correlated anisotropic rough surfaces defined by the correlation lengths $a_2=1.5\lambda$ [and $a_1=\lambda$]. From the contour plot of the surface realization of such surfaces presented in Fig.~\ref{Fig:AnisotropicSurfaces-SurfaceRoughness}(b), it is not immediately apparent that the surface is anisotropic; to realize this, a close inspection of the topography is needed, and as a result, we will in the following refer to this class of surfaces as being \emph{slightly} anisotropic. Figures~\ref{Fig:AnisotropicSurfaces}(d)--(f) display the angular dependence of the mean DRCs for the azimuthal angles of incidence  $\phi_0=\ang{0}$, \ang{45} and \ang{90}, respectively. The first thing to notice from these results is that the anisotropy of the surface roughness alters the scattered intensity distributions relative to the intensity distributions  obtained for the corresponding isotropic surface for which $a_2=\lambda$~[Figs.\ref{Fig:AnisotropicSurfaces}(a)--(c)]. It is also found that the mean DRCs for this anisotropic surface have their highest scattered intensity at, or close to, the backscattering (retroreflection) direction~[black crosses in Fig.~\ref{Fig:AnisotropicSurfaces}]. However, the angular dependencies of the scattered intensity around these directions are different for the isotropic and anisotropic surfaces. For instance, by comparing the mean DRCs in Figs.~\ref{Fig:AnisotropicSurfaces}(d) and \ref{Fig:AnisotropicSurfaces}(a) it is observed that what appears as an almost isotropic feature around the backscattering direction in the isotropic case~[Fig.~\ref{Fig:AnisotropicSurfaces}(a)], is transformed into an elliptic-like intensity distribution~[Fig.~\ref{Fig:AnisotropicSurfaces}(d)] about the same direction that is elongated along the $\vecUnit{q}_2$ direction, that is, along the direction for which the correlation length is the longest. At the same time, the scattered intensity in the backscattering direction is lower in the anisotropic case than in the isotropic case.

A similar kind of elongation of the scattered intensity distribution is observed for the anisotropic case when $\phi_0=\ang{90}$~[Fig.~\ref{Fig:AnisotropicSurfaces}(f)]; however, in this case the elongation is predominantly in a positive $\vecUnit{q}_2$-direction away from the retroreflection direction and extending all the way to the specular direction. A comparison of the mean DRCs in Figs.~\ref{Fig:AnisotropicSurfaces}(f) and ~\ref{Fig:AnisotropicSurfaces}(c) also reveals that the anisotropy of the surface enhances the scattering into the forward scattering plane, that is, into directions for which $q_2>0$. It ought to be remarked that in the scattering from isotropic surfaces we also observed that increasing the correlation length $a>\lambda$ caused broadening of the mean DRCs as well as enhanced scattering into the forward scattering plane; see Figs.~\ref{Fig:Dirichlet_cuts_rms100} and \ref{Fig:Dirichlet_rms100}.

Figure~\ref{Fig:AnisotropicSurfaces}(e) presents the mean DRC for an anisotropic surface for which $a_2=1.5\lambda$ and $\phi_0=\ang{45}$. In this case we also observe  an elongation and shift of the scattered intensity distribution around the retroreflection direction into direction $\vecUnit{q}_2$. The red elliptic-like  structures seen in Figs.~\ref{Fig:AnisotropicSurfaces}(e)--(f) are rather similar. Moreover, also for the case when $\phi_0=\ang{45}$~[Fig.~\ref{Fig:AnisotropicSurfaces}(e)] an enhancement of the scattering into the forward scattering plane is observed; this is quite similar to what was found previously when $\phi_0=\ang{90}$~[Fig.~\ref{Fig:AnisotropicSurfaces}(f)].

The mean DRCs for the isotropic surface shown in  Figs.~\ref{Fig:AnisotropicSurfaces}(a)--(c) all have their highest intensity in the backscattering direction. For the anisotropic surface of correlation length $a_2=\num{1.5}\lambda$~[Figs.~\ref{Fig:AnisotropicSurfaces}(d)--(f)] we find that this is also the case when $\phi_0=\ang{0}$~[Fig.~\ref{Fig:AnisotropicSurfaces}(d)]. However, for the same roughness parameters and when $\phi_0=\ang{45}$ and \ang{90}, a close inspection of the mean DRCs in Figs.~\ref{Fig:AnisotropicSurfaces}(e)--(f) reveals that this is no longer the case; here the directions of maximum intensity are instead shifted away from the retroreflection direction by a small amount in the positive $\vecUnit{q}_2$-direction. Such shifts of the maximum diffusely scattered intensity are even more apparent when the correlation length $a_2$ is increased further.
Figures ~\ref{Fig:AnisotropicSurfaces}(g)--(i) and \ref{Fig:AnisotropicSurfaces}(j)--(l) present the mean DRCs corresponding to the correlation lengths $a_2=2\lambda$ and $a_2=3\lambda$, respectively, when the other roughness and numerical parameters are the same. For both these values of $a_2$, and when $\phi_0=\ang{45}$ and \ang{90}, one finds that the maxima of the scattered intensity are found in the region $q_2>0$ which is far from the value of the 2nd wave vector coordinate of the backscattering direction. On the other hand, the 1st coordinate of the points of maximum intensity and the backscattering direction seems to be only little affected, if at all, by the anisotropy of the surface. In particular, when $\phi_0=\ang{90}$ we find from Figs.~\ref{Fig:AnisotropicSurfaces}(i,\,l) that the direction of maximum scattered intensity is at, or close to, the specular direction. This we speculate is caused by the wave scattering from the grooves of the surface roughness~[see Figs.~\ref{Fig:AnisotropicSurfaces-SurfaceRoughness}(c)--(d)]; such scattering will cause most of the intensity being scattered into the forward direction when $a_2>\lambda$, similarly to what we saw previously for the scattering from isotropic surfaces. Furthermore, when $\phi_0=\ang{0}$, we observe from the results in Figs.~\ref{Fig:AnisotropicSurfaces}(g,\,j) that the highest scattered intensity remains located in the backscattering direction also for these strongly anisotropic surface; it is speculated that this will be true when $\phi_0=\ang{0}$ for any value of $a_2$ assuming  $a_1=\lambda$ and $\delta=\lambda/2$.

By comparing the angular dependence of the mean DRCs for the correlation lengths  $a_2=2\lambda$~[Figs.~\ref{Fig:AnisotropicSurfaces}(g)--(i)] and
$a_2=3\lambda$~[Figs.~\ref{Fig:AnisotropicSurfaces}(j)--(l)], several observations can be make. First, the scattered intensity distributions are found to be significantly narrower in the $\vecUnit{q}_2$-direction when $a_2=3\lambda$ than what it is when $a_2=2\lambda$. On the other hand, the widths of the same distributions in the $\vecUnit{q}_1$-direction are only very little affected by the change in the $a_2$ correlation length. This behavior we attribute to the reduced height difference over a wavelength the surface has in the $x_2$ direction when the correlation length $a_2$ is several times larger than the wavelength $\lambda$. In the extreme limit  that $a_2\rightarrow\infty$, the mean DRC for $\phi_0=\ang{0}$ will be proportional to $\delta(q_2)$, which has zero width in the $\vecUnit{q}_2$ direction, and it is the transition towards this limit that can be observed in the first column of Fig.~\ref{Fig:AnisotropicSurfaces}. Second, the angular dependence of the scattered intensity about the direction of highest scattered intensity is significantly more isotropic for the case when  $a_2=3\lambda$~[Figs.~\ref{Fig:AnisotropicSurfaces-SurfaceRoughness}(j)--(l)] than what it is for the shorter correlation length $a_2=2\lambda$~[Figs.~\ref{Fig:AnisotropicSurfaces-SurfaceRoughness}(g)--(i)].
In this regard, the situation seen when $a_2=3\lambda$ resembles more the case of the isotropic surface in Figs.~\ref{Fig:AnisotropicSurfaces}(a)--(c); however, the directions of maximum scattered intensity are rather different in these two cases.
Finally, when  $\phi_0=\ang{45}$ it is readily observed from the results in  Figs.~\ref{Fig:AnisotropicSurfaces}(h,\,k) that the directions where the mean DRCs have their maxima are not in the plane of incidence. This possibility was already alluded to earlier when discussing the result in Fig.~\ref{Fig:AnisotropicSurfaces}(e). However, what determines the direction of the highest scattered intensity? A hint towards an explanation is found by observing from the results in Fig.~\ref{Fig:AnisotropicSurfaces} that the $q_1$ coordinate of the directions of the highest scattered intensity is only marginally affected, if at all, by the anisotropy of the surface introduced by increasing the correlation length $a_2$ to values larger then $\lambda$. Moreover, when the correlation length $a_2$ is significantly larger then $a_1=\lambda$, the incident beam will scatter in a more specular-like fashion from the ridges and grooves of the strongly anisotropic surface. For the roughness parameters that we assume, the maxima of the mean DRCs for an isotropic surface are in the backscattering directions defined by $\pvec{q}=-\pvec{k}=(-k_1,-k_2,0)$. The heuristic argument presented above for a strongly anisotropic surface with $a_2>a_1\geq \lambda$, predicts that the maximum  of the mean DRC will be in the direction $\pvec{q}=(-k_1,k_2,0)$; This direction is indicated by black triangles in the 2nd column in Fig.~\ref{Fig:AnisotropicSurfaces}.
For instance, this prediction agrees rather well with what is observed in the computer simulation results for the anisotropic surface with correlation lengths $a_1=\lambda$ and $a_2=3\lambda$ that are presented in Figs.~\ref{Fig:AnisotropicSurfaces}(j)--(l). A rigorous derivation of the above result for these and other roughness parameters, we will leave for a dedicated study.

%
%
One final observation should be made about the angular intensity distributions depicted in Fig.~\ref{Fig:AnisotropicSurfaces}. When the plane of incidence contains one of the two main axes of the anisotropy, in our case, $\vecUnit{q}_1$ or $\vecUnit{q}_2$, then the mean DRCs are expected to display a \emph{reflection symmetry} with respect to the plane of incidence. That this is the case can be observed from the mean DRCs presented in the first and last columns of  Fig.~\ref{Fig:AnisotropicSurfaces}; the error in the satisfaction of the reflection symmetry in these results we ascribe to the use of a finite number of surface realization in calculating these mean DRCs. When $\phi_0$ does not equal \ang{0} or \ang{90} (and $\ang{0}\leq \phi_0\leq \ang{90}$), this reflection symmetry with respect to the plane of incidence is not expected to hold, something that can be observed from the panels forming the 2nd column of Fig.~\ref{Fig:AnisotropicSurfaces}.

\begin{figure*}[tbp!]
  \centering
  \includegraphics[width=\textwidth]{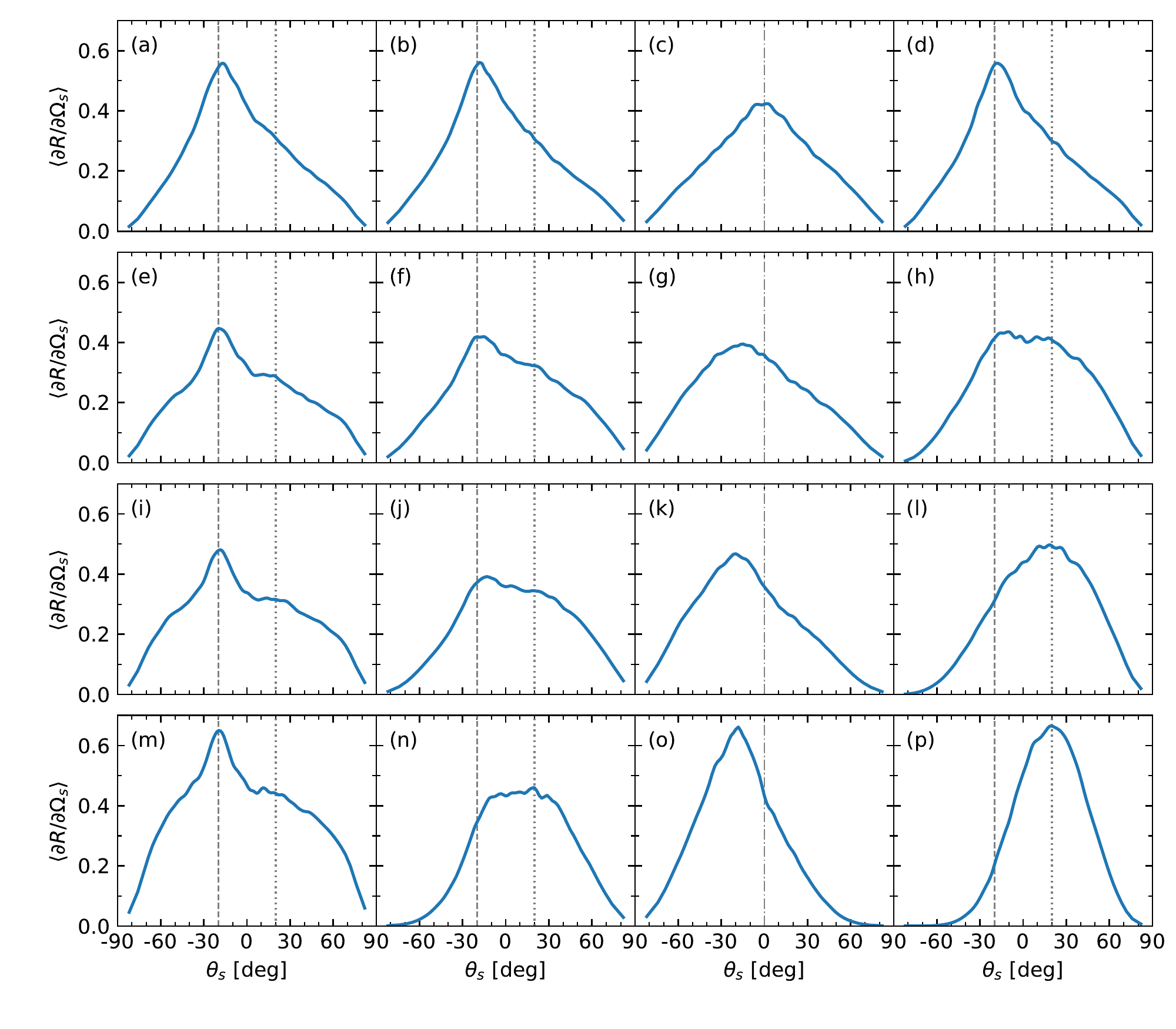}
  \caption{Various in-plane and out-of-plane behaviors with the polar scattering angle $\theta_s$ for the angular dependent mean DRCs presented in Fig.~\protect\ref{Fig:AnisotropicSurfaces}. Out-of-plane plots are presented in the 3rd column while the plots in the other columns are in-plane plots. The polar angle of incidence assumed is $\theta_0=\ang{20}$, while the azimuthal angle of incidence is $\phi_0=\ang{0}$~[column~1]; $\phi_0=\ang{45}$~[columns~2 and 3]; and  $\phi_0=\ang{45}$~[column~4]. The results that appear in a given row of this figure were obtained from the mean DRCs appearing in the same row of Fig.~\protect\ref{Fig:AnisotropicSurfaces}. Therefore the surface roughness parameters are $a_2=\lambda$~[1st row]; $a_2=\num{1.5}\lambda$~[2nd row]; $a_2=2\lambda$~[3rd row]; $a_2=3\lambda$~[4th row], and in all cases $\delta=\lambda/2$ and $a_1=\lambda$. The remaining numerical parameters are identical to those of Fig.~\protect\ref{Fig:AnisotropicSurfaces}. The vertical lines mark the positions of the backscattering direction~(dashed lines), specular direction~(dotted lines) and the $\theta_s=\ang{0}$ direction~(dash-dotted lines).} 
  \label{Fig:AnisotropicSurfaces-cuts}
\end{figure*}

\smallskip
Until now not much attention have been given to the amplitudes of the mean DRCs for anisotropic surfaces. To address this and other issues, in Fig.~\ref{Fig:AnisotropicSurfaces-cuts} we present the in-plane and out-of-plane angular dependencies of the mean DRCs from Fig.~\ref{Fig:AnisotropicSurfaces}. The panels of this figure show the in-plane angular dependencies, except for the 3rd column that depicts the out-of-plane dependence. The azimuthal angles of incidence are $\phi_0=\ang{0}$~[column~1]; $\phi_0=\ang{45}$~[columns~2 and 3]; and $\phi_0=\ang{90}$~[column~4] while for all cases the polar angle of incidence is $\theta_0=\ang{20}$. For the azimuthal angle of incidence $\phi_0=\ang{0}$, the amplitudes of the in-plane mean DRCs do depend on the level of anisotropy. It is observed from the results in column~1 of Fig.~\ref{Fig:AnisotropicSurfaces} that the amplitude of the in-plane mean DRCs initially drops with increasing value of $a_2$ before it starts increasing again when the same parameter is increased further. Such behavior can be understood in terms of the observation we did previously regarding the width of the angular dependencies of the mean DRC in the out-of-plane directions when discussing Fig.~\ref{Fig:AnisotropicSurfaces};  the widths of these distributions initially increase with increasing $a_2$ before they gradually decrease with the increase of the same parameters as the full angular distributions of the mean DRCs become centered around the plane of incidence. Note that for $\phi_0=\ang{0}$ there are well pronounced backscattering peaks present in all the in-plane dependencies of the mean DRCs presented in column~1 of Fig.~\ref{Fig:AnisotropicSurfaces-cuts}.

When the azimuthal angle of incidence is $\phi_0=\ang{90}$, column~4 of Fig.~\ref{Fig:AnisotropicSurfaces-cuts}, the dependence of the amplitudes of the in-plane angular distributions on the correlation length $a_2$ is found to be similar to what was found when   $\phi_0=\ang{0}$. However, there are also significant differences between the two cases. From column~4 of Fig.~\ref{Fig:AnisotropicSurfaces-cuts} it is rather apparent how an increase from unity of the anisotropy ratio $a_2/a_1$ causes a shift of the maxima of the in-plane scattered intensity distributions from the backscattering direction~[Fig.~\ref{Fig:AnisotropicSurfaces-cuts}(d)] to the specular direction~[Fig.~\ref{Fig:AnisotropicSurfaces-cuts}(p)]. For instance, when $a_2/a_1=3$ the in-plane dependence of the mean DRC~[Fig.~\ref{Fig:AnisotropicSurfaces-cuts}(p)] is already well centered around the specular direction. The reason for this behavior is explained in our discussion of Fig.~\ref{Fig:AnisotropicSurfaces}.

We now turn to the situation for which $\phi_0=\ang{45}$ presented in columns~2 and 3 of Fig.~\ref{Fig:AnisotropicSurfaces-cuts} for in-plane and out-of-plane distributions, respectively. The amplitudes of the in-plane results for this azimuthal angle of incidence, seem not to increase with the value of $a_2$ as we saw previously for the cases $\phi_0=\ang{0}$ and \ang{90}. This is partly caused by the maxima of the mean DRC moving out of the plane of incidence, see Fig.~\ref{Fig:AnisotropicSurfaces}. However, the most interesting results for $\phi_0=\ang{45}$ are found for the out-of-plane distributions~[Fig.~\ref{Fig:AnisotropicSurfaces-cuts}, column~3]. It is found that the reflection symmetry of this distribution corresponding to an isotropic surface is lost as the anisotropy ratio $a_2/a_1$ is increased from unity. For the parameters that we have assumed the maxima of these out-of-plane distributions are found for $\theta_s<\ang{0}$ and their amplitudes are found to increase with increasing $a_2/a_1$. 

\smallskip
It should be mentioned that we also did simulations for less rough anisotropic surfaces for which $\delta=\lambda/10$, as in Fig.~\ref{Fig:Reflectivity}, and assuming the same correlation lengths and polar angle of incidence as were used in producing the results in Figs.~\ref{Fig:AnisotropicSurfaces}  and \ref{Fig:AnisotropicSurfaces-cuts}. The aim was to investigate the dependence of the reflectivity on the azimuthal angle of incidence $\phi_0$. It was found that the variation of the reflectivity with $\phi_0$ was of the order $\num{E-4}$, which  is too small to be significant with the precision that we have in our simulations, at least, this was the case for the roughness and numerical parameters that we assumed in performing them.

\section{\label{sec:conclusion}Conclusion}

In conclusion, we have by  numerical simulations studied the scattering of incident Gaussian scalar beams from isotropic and anisotropic, strongly rough Dirichlet and Neumann surfaces. To this end, we present the rigorous, inhomogenious integral equations for the field~(Neumann surface) or the normal derivatives of the field~(Dirichlet surface) that are obtained by the use of the Green's function surface integral method. By a nonperturbative and purely numerical solution of these integral equations, we obtain the fields scattered from the rough surfaces. For a set of roughness parameters and angles of incidence, we calculate the full angular distribution of the mean DRCs for isotropic, Gaussian correlated  Dirichlet and Neumann surfaces. The surface parameters were chosen so that some of the results showed enhanced backscattering peaks, which is the hallmark of multiple scattering processes.  The quality of the simulation results was quantified by investigating energy conservation~(unitarity), and it was found to be satisfied with an error smaller than \num{2E-4}, or better, for the main results presented.  We also calculated the dependence of the reflectivity on the polar angle of incidence. It was found that for the same parameters characterizing the isotropic rough surface, the reflectivity of a Dirichlet surface was always higher then the reflectivity of the corresponding Neumann surface independent of the polar angle of incidence. 

For anisotropic, Gaussian correlated, strongly rough surfaces we calculated the full angular distribution of the mean DRCs for both Dirichlet and Neumann surfaces for a given polar angle of incidence [$\theta_0=\ang{20}$] and three azimuthal angles of incidence $\phi_0=\ang{0}$, \ang{45}, or \ang{90}. We found  that even for moderate levels of surface anisotropy, as characterized by the ratio of the two correlation lengths, $a_2/a_1$, the full angular distributions of the scattered intensity (mean DRCs) were affected in a profound manner by the surface anisotropy. For a set of four surface roughness parameters, corresponding to an increasing ratio of surface anisotropy, we calculated and presented the full angular distribution of the mean DRCs. The features of the scattered intensity distributions obtained in this way were discussed and rationalized.

\begin{acknowledgments}
The work of T.N. received support from the Research Council of Norway, Fripro Project No.~213453. The research of I.S. was supported in part by the Research Council of Norway~(Contract 216699) and the French National Research Agency (ANR-15-CHIN-0003). This research was supported in part by NTNU and the Norwegian metacenter for High Performance Computing (NOTUR) by the allocation of computer time.

T.S.H. and T.N. contributed equally to this work.
\end{acknowledgments}

%

%

\end{document}